\documentclass[aps,prapplied,twocolumn,amsmath,amsthm,amssymb,showpacs,floatfix,superscriptaddress,longbibliography]{revtex4-2}
\usepackage[utf8]{inputenc}
\usepackage{dsfont}
\usepackage{xcolor}
\usepackage{float}
\usepackage{enumitem}
\usepackage{amsmath}
\usepackage[normalem]{ulem}
\usepackage{siunitx}
\usepackage{graphicx}
\usepackage{adjustbox}
\usepackage{dcolumn}
\usepackage{bm}
\usepackage[unicode=true]{hyperref}
\hypersetup{colorlinks=true,linkcolor=blue,filecolor=blue,urlcolor=blue,citecolor=blue}
\usepackage{nicefrac}
\usepackage{physics}
\usepackage[capitalize]{cleveref}
\usepackage[caption=false]{subfig}
\newcommand{\phantomsubfloat}[1]{
	{
		\captionsetup[subfigure]{labelformat=empty}
		\subfloat[][]{#1}
	}%
}
\usepackage{tikz}
\usetikzlibrary{arrows}
\usepackage{listings}
\newcommand{\cOne}[0]{{$\textrm{C}_1$}}
\newcommand{\cTwo}[0]{{$\textrm{C}_2$}}
\newcommand{\cThree}[0]{{$\textrm{C}_3$}}
\newcommand{\CCC}[0]{$\mathds{C}^3$}
\newcommand{\EE}[2]{#1\times 10^{#2}}

\begin{document}

\title{
Integrated tool-set for control, calibration, and characterization of quantum devices applied to superconducting qubits
}


\author{Nicolas Wittler}
\thanks{These two authors contributed equally}
\email{c3@q-optimize.org}
\affiliation{Theoretical Physics, Saarland University, 66123 Saarbr\"ucken, Germany}
\affiliation{Peter Gr\"unberg Institut -- Quantum Computing Analytics (PGI 12), Forschungszentrum J\"ulich, D-52425 J\"ulich, Germany}

\author{Federico Roy}
\thanks{These two authors contributed equally}
\email{c3@q-optimize.org}
\affiliation{Theoretical Physics, Saarland University, 66123 Saarbr\"ucken, Germany}
\affiliation{IBM Quantum, IBM Research GmbH, Zurich Research Laboratory, S\"aumerstrasse 4, 8803 R\"uschlikon, Switzerland}

\author{Kevin Pack}
\affiliation{Theoretical Physics, Saarland University, 66123 Saarbr\"ucken, Germany}
\affiliation{Peter Gr\"unberg Institut -- Quantum Computing Analytics (PGI 12), Forschungszentrum J\"ulich, D-52425 J\"ulich, Germany}

\author{Max Werninghaus}
\affiliation{Theoretical Physics, Saarland University, 66123 Saarbr\"ucken, Germany}
\affiliation{IBM Quantum, IBM Research GmbH, Zurich Research Laboratory, S\"aumerstrasse 4, 8803 R\"uschlikon, Switzerland}

\author{Anurag Saha Roy}
\affiliation{Theoretical Physics, Saarland University, 66123 Saarbr\"ucken, Germany}

\author{Daniel J. Egger}
\affiliation{IBM Quantum, IBM Research GmbH, Zurich Research Laboratory, S\"aumerstrasse 4, 8803 R\"uschlikon, Switzerland}

\author{Stefan Filipp}
\affiliation{IBM Quantum, IBM Research GmbH, Zurich Research Laboratory, S\"aumerstrasse 4, 8803 R\"uschlikon, Switzerland}
\affiliation{Department of Physics, Technical University of Munich, 85748 Garching, Germany}
\affiliation{Walther-Meißner-Institut, Bayerische Akademie der Wissenschaften, Garching 85748, Germany}

\author{Frank K. Wilhelm}
\affiliation{Theoretical Physics, Saarland University, 66123 Saarbr\"ucken, Germany}
\affiliation{Peter Gr\"unberg Institut -- Quantum Computing Analytics (PGI 12), Forschungszentrum J\"ulich, D-52425 J\"ulich, Germany}

\author{Shai Machnes}
\affiliation{Theoretical Physics, Saarland University, 66123 Saarbr\"ucken, Germany}
\affiliation{Peter Gr\"unberg Institut -- Quantum Computing Analytics (PGI 12), Forschungszentrum J\"ulich, D-52425 J\"ulich, Germany}

%
\begin{abstract}
Efforts to scale-up quantum computation have reached a point where the principal limiting factor is
not the number of qubits, but the entangling gate infidelity.
However, the highly detailed system characterization required to understand the underlying error sources is an arduous process and impractical with increasing chip size.
Open-loop optimal control techniques allow for the improvement of gates but are limited by the models 
they are based on.
To rectify the situation, we provide an integrated open-source tool-set for Control, Calibration and Characterization
(\CCC), capable of open-loop pulse optimization, model-free calibration, model fitting and refinement.
We present a methodology to combine these tools to find a quantitatively accurate system model,
high-fidelity gates and an approximate error budget, all based on a high-performance, feature-rich 
simulator.
We illustrate our methods using simulated fixed-frequency superconducting qubits for which we learn model parameters with less than $1\%$ error and derive a coherence limited cross-resonance (CR) gate that achieves $99.6\%$ fidelity without need for calibration.
\end{abstract}

\maketitle


\section{The problem}\label{sec:problem}

Scaling up quantum processing units (QPUs) is a monumental task, that requires the community to make
progress on multiple fronts, most importantly improving gate fidelities and increasing the number of qubits. Over the past few years, significant emphasis has been placed on creating larger devices, yielding great success \cite{jurcevic2020demonstration, arute2019quantum}.
However, the number of qubits has outstripped the limits that fidelity places on their utility:
In \cite{jurcevic2020demonstration}, a record quantum volume \cite{cross2019validating} of 64 was demonstrated, loosely translating to the device being able to perform $\log_2\left(64\right)^2 = 36$ entangling gates before fidelity drops below $\nicefrac{2}{3}$, a relatively small number of gates for an array of six qubits;
In \cite{arute2019quantum} the circuit fidelity was $0.1\%$ thus requiring 30 million repetitions to achieve the desired statistics.
One could even argue that the two-qubit gate fidelities demonstrated in isolation in 2014 \cite{Barends2014} ($0.994$) are comparable with those in 2019's
\cite{arute2019quantum} ($0.9938$), even though the latter are for simultaneous gates in a large 2D qubit array.

The relatively slow progress in improving gate fidelities can be traced back to an incomplete
understanding of the sources of error. Indeed, characterization and calibration of QPUs to the
desired accuracy is impractical and cumbersome, and operating on devices of increasing qubit number requires entangling gates to be fine-tuned for each individual pair to account for slightly varying properties.
The resulting lack of detailed models makes it
harder to identify where efforts must be focused to achieve higher fidelity gates  \cite{collodo2020implementation, ganzhorn2020benchmarking}.

Given that ``all models are wrong, but some are useful'' \cite{box1976science}, we describe a Good Model as follows:\\
\vspace{-14pt}
\begin{center}
    \begin{minipage}{0.43\textwidth}
        \begin{center}
            A \emph{Good Model} is \hypertarget{def-goodmodel}{one that predicts the behavior of
            the system, for the operations we wish to perform, to accuracies we care about}.
        \end{center}
    \end{minipage}
\end{center}
\vspace{-6pt}
For a QPU, a Good Model has to have predictive power for the range of feasible
gate-generating pulses and for long sequences of such gates, to a fidelity accuracy of the order of
$10^{-5}$. To the authors' knowledge, no such Good Model for a superconducting QPU has ever been
published.

Since models serve as the basis to derive high-fidelity gates in open-loop optimal control 
\cite{Glaser2015, Machnes2011, Machnes2018, GRAPE, CRAB, Krotov1, DRAG, Schutjens2013}, any inaccuracies of the model will inevitably degrade the experimental accuracy of the resulting gates.
This problem is only partially ameliorated by the first-order insensitivity of optimized pulses to model inaccuracies \cite{dorner2005quantum, Egger2013}.
Methodologies such as the adaptive hybrid optimal control (Ad-HOC) protocol \cite{Egger2014} -- which combines a model-based open-loop optimization with a closed-loop experimental calibration \cite{CRAB, ORBIT} -- address this issue but leave one in an unsatisfactory position as the need for calibration proves the inadequacy of the model: the root causes of the remaining infidelities are unexplained.

Conversely, if a Good Model is known, gates generated by open-loop optimal control will, by
definition, work on the experiment, not requiring further closed-loop calibration. This enables
the use of complex pulses that would otherwise require time-consuming calibration.
Such a Good Model
would also provide an error budget through a process of exploratory
interrogation -- evaluating the potential performance of the system where certain limitations
have been removed, i.e. asking ``what if ...?''.
Therefore, extracting a Good Model efficiently and in a highly automated manner is key to
improving fidelities and a crucial step of QPU scale-up.

In this work we present \CCC, our proposed approach to control, calibrate and characterize QPUs.
The paper is organized as follows: We present the conceptual steps of \CCC\ in Sec. \ref{sec:CCC} and illustrate the methodology by example in Sec. \ref{sec:example}, showing how these steps are implemented. Sec. \ref{sec:CCC-in-detail} includes a detailed description of the modeling, optimization procedures, the data comparison function and relevant prior work.
We conclude in \ref{sec:discussion} with a discussion of the effort's current status and long-term directions.


\section{\CCC\ -- Control, Calibration, and Characterization}\label{sec:CCC}

\begin{figure}[b]
	\includegraphics[width=\linewidth]{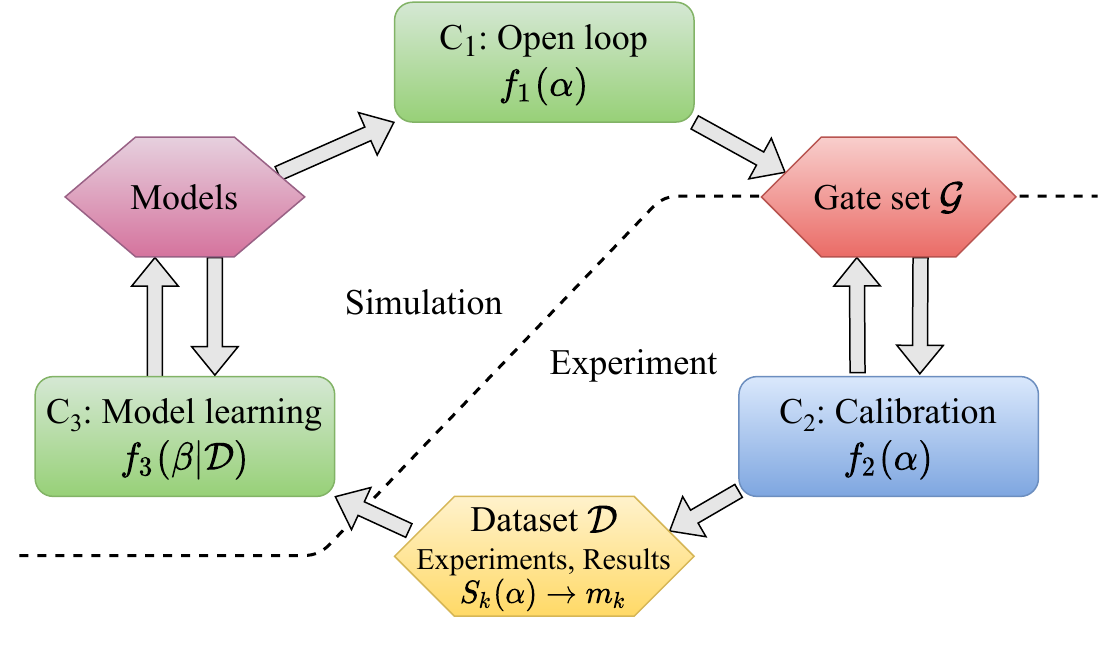}
	\caption{
		Diagram of the \CCC\ tool-set in an integrated characterization loop.
		\textbf{\cOne}\ is a tool for obtaining optimal pulses by finding the control parameters $\alpha$
		that minimize a goal function $f_1(\alpha)$ in simulation.
		The \textbf{gate-set} $\mathcal{G}$ includes all the operation that one wishes to perform on the
		experiment, including the information of the ideal logical operations and the optimal 
		pulse parameters $\alpha$ that implement them.
		\textbf{\cTwo}\ is a model-free experimental calibration procedure that optimizes pulse shapes with
		a gradient-free search to minimize an infidelity function $f_2(\alpha)$ by varying all
		parameters at once.
		A \textbf{data-set} is a collection of experiment/result pairs, including information about the pulses
		parameters used $\alpha$, the sequences $S_k$ performed and the final outcomes measured, $m_k$.
		\textbf{\cThree}\ is a tool for model learning that determines the model parameters $\beta$ that best
		explain the data-set. It minimizes a goal function $f_3(\beta\vert \mathcal{D})$ obtained by
		recreating experiments $S_k(\alpha)$ in simulation and comparing the results to the ones in the experiment.
		In \CCC\ different parameterized models can be provided to represent various elements of the 
		experiment to find the one that best describes it.
		After the learning, the resulting model can be the basis for another characterization loop, refining both
		model and controls.
	}
	\label{fig:c3po}
\end{figure}

Current methodology relies on tailored routines to extract individual parameters of the system's
model (characterization) or fine-tune specific parameters of pulses used (calibration) \cite{Sheldon2016, Sheldon2016single}. 
This approach becomes cumbersome and impractical as the number of model and pulse parameters increases.
With \CCC\ we propose a different paradigm: Optimizing a figure of merit that is sensitive to the set of parameters we care about.
This eliminates the need to design per-parameter measurements, and thus provides a more general approach.
\CCC\ at its core is composed of three separate optimizations, respectively implementing the tasks of control, calibration, and characterization:
\begin{description}[leftmargin=8pt]
    \item[\cOne] Given a model, find the pulse shapes that maximize fidelity with a target operation. Pulse shapes may be constrained by an ansatz or allow direct arbitrary waveform
         generator (AWG) parameterization.
    \item[\cTwo] Given pulse shapes, calibrate their parameters, if possible simultaneously, to maximize a figure of merit measured by the actual
         experiment, thus improving beyond the limits of a deficient model.
    \item[\cThree] Given control pulses and their experimental measurement outcome, optimize model parameters to best reproduce the results. Enhance
         the model if needed.
\end{description}

The tasks of open-loop optimal control, \cOne, and calibration, \cTwo, are fairly established in the community
\cite{Glaser2015,Machnes2011,Machnes2018,GRAPE, CRAB, Krotov1, DRAG, Schutjens2013, Egger2014,CRAB,ORBIT}.
To characterize the system and provide us with a Good Model, we introduce \cThree, a tool to optimize model parameters by comparing model prediction to experimental data. We refer to this task as model learning.
For this purpose, one requires an experimental data-set containing information about the implemented pulses and the corresponding measurement outcomes. 
To test the model accuracy, we reproduce the data-set, applying the same pulses to a simulation of the experiment, and compare the resulting outcomes: This provides a model match score to optimize. 
Initially, a candidate model is formulated based on previous information or intuition. 
If the model is suitable to explain the experiment, the optimization will converge to a near perfect 
match, thus providing numeric values for the model parameters.
Instead, if the match is poor, the user supplies a new model, that is either an extension or modification of the previous candidate, and the optimization is repeated.
Depending on user choice, learned values are carried over to the parameters of the new model or discarded.

As heterogeneous experimental data is the foundation for model learning, we suggest using the three tasks of
\CCC\ in sequence, as shown in Fig. \ref{fig:c3po}. However,  their application is by no means
limited to this use case and one may choose to view them as stand-alone routines. 
The same tools used to realize \cOne, \cTwo\ and \cThree\ can also be used to further interrogate
the system to obtain a sensitivity analysis of the optimized model in light of the
experimental data and a breakdown of possible error sources.

We note that the intertwining of control and characterization has been raised in the
more general context of control theory \cite{geversaa2005identification,rojas2007robust,ljung2010perspectives,bombois2011optimal}.
In quantum technology, there are some works which combine two of the three
tasks: Ad-HOC \cite{Egger2014} calls for optimal control followed by
calibration; a combination of model-based gradient calculations and
experimental calibrations is proposed by \cite{chen2018combining,wu2018data},
but the data gathered is not used to improve the system model;
in \cite{ball2020software} pulses are designed specifically for the purpose of reconstructing the noise spectrum.


\section{Synthetic Application Example}\label{sec:example}

The following synthetic example illustrates how \CCC\ is used to obtain a Good Model in a realistic setting.
We simulate a two-qubit QPU device using an underlying model, labeled the ``real'' model,
which includes control discretization effects, electronics transfer functions, Markovian noise, and state preparation and measurement (SPAM) errors.

In this example, the simulated device is treated as a black-box, which we interrogate with \CCC. We derive (\cOne) and calibrate (\cTwo) optimal control pulses and use the resulting data to extract a Good Model (\cThree) by comparing the black-box to three candidates:
\begin{description}
  \item[Simple model] Two \emph{uncoupled} qubits, \emph{closed} system dynamics;
  \item[Intermediate model] Two \emph{coupled} qubits, \emph{closed} system dynamics;
  \item[Full model] Two \emph{coupled} qubits, \emph{open} system dynamics, including \emph{SPAM} errors. 
  Same structure as the ``real'' model, but \emph{a priori} undetermined parameter values.
\end{description}
We systematically enrich the model until it reproduces the behavior of the device observed in \cTwo. The recovered model is then used to design a two-qubit gate that performs well on the black-box without the need for tune-up.

\subsection{The black-box device (``real'' model)} 

The ``real'' model is composed of two coupled three-level Duffing oscillators, labeled by $A$ and $B$, 
each directly driven by an external field $c_i(t)$. Initialization, dynamics and readout are performed in the
dressed basis by solving the master equation in Lindblad form \cite{lindblad1976generators,gorini1976completely}.
\begin{equation}\label{eq:lindblad-master}
    \begin{aligned}
    \dot{\rho} = -i [H, \rho] + \sum_{\substack{i=A,B \\ j=\phi,\kappa}} L_{i,j} \rho L_{i,j}^\dagger- \frac{1}{2}
                                \qty{L_{i,j} L_{i,j}^\dagger, \rho}
    \end{aligned}
\end{equation}
with
\begin{equation}
    \begin{aligned}
        H/\hbar =& \sum_{i=A,B} \left[
                        \omega_{i}b_i^\dagger b_i - \frac{\delta_i}{2}
                        \qty(b_i^\dagger b_i - 1) b_i^\dagger b_i 
                    \right] +\\ 
                +& g(b_A+b_A^\dagger)(b_B+b_B^\dagger)
                    + \sum_{i=A,B}   c_i(t) \qty(b_i+b_i^\dagger) \ ,
    \end{aligned}
\end{equation}
where $\omega_{i}$ is the frequency of qubit $i$, $\delta_i$ is the anharmonicity, $b_i$
($b_i^\dagger$) is the lowering (raising) operator, and $g$ is the coupling strength.
Open-system effects are expressed by the dephasing and relaxation Lindblad operators $L_{i,\phi} =\sqrt\frac{2}{T_i^{2*}} b_i b_i^\dagger $ and $L_{i,\kappa} = \sqrt\frac{1}{T^1_i}b_i$ with decay rates $1/T^1_i $ and $1/T_i^{2*}$.\\
Given the input drive signals $\varepsilon_i(t)$, we calculate the effective control fields $c_i(t) =
\varphi[\varepsilon_i(t)]$, where the transfer function $\varphi$ \cite{girod2001signals} accounts
for discretization introduced by the AWG, bandwidth limitations of hardware, and for a constant scaling
$\varphi_0$, which translates input voltages to field amplitudes. 
We implement state preparation errors due to a non-zero initial temperature $T$ by starting each 
experiment from the thermal state
\begin{equation}\label{eq:thermal}
    \begin{aligned}
    \rho_\text{init}=\frac{1}{Z} 
    & \left[\ketbra{0} +\exp{-\frac{E_1}{k_BT}}\ketbra{1}\right. \\
    &\left. +\exp{-\frac{E_2}{k_BT}}\ketbra{2}\right]
    \end{aligned}
\end{equation}
where $Z=\sum_{k=0}^2\exp{-E_k / k_B T}$ is the partition function with energies $E_{0,1,2} = 0,\hbar\omega_q,\hbar(2\omega_q+\delta)$, 
and $k_B$ is the Boltzmann constant.
readout misclassification is included, measuring state $\ket{n}$ as state $\ket{m}$ with
probability $p_{n \rightarrow m}$. For example the probability of measuring a state $\rho_\psi=\ketbra{\psi}$ as $\ketbra{0}$ is
\begin{equation}\label{eq:readout}
    \begin{aligned}
        \Pi_0(\rho_\psi)=& ~p_{0 \rightarrow 0} ~ \mel{0}{\rho_\psi}{0}
               + ~p_{1 \rightarrow 0}  ~ \mel{1}{\rho_\psi}{1}\\
               &+ ~p_{2 \rightarrow 0}  ~ \mel{2}{\rho_\psi}{2}.
    \end{aligned}
\end{equation}
Similarly to experiment, populations are estimated by averaging the results of multiple projective measurements,
simulated as a multinomial draw from the distribution with probabilities $\qty{\Pi_n}$, thus introducing noise stemming from a
finite number of measurement repetitions (commonly known as `shot noise').
The values of the ``real'' model parameters are summarized in Tab. \ref{tab:parameters}.

\begin{table*}[t]
	\caption{Overview of the parameters of the ``real'' model (reference values), and the candidate models, before and after the \cThree\ learning for different data-sets. Candidate model values are shown as difference from reference values.
		Dashes (--) indicate parameters not present in the model, quotation marks (") indicate parameters not being changed.
	}
	\footnotesize
	\centering
	\def\arraystretch{1.5}
	\setlength\tabcolsep{0.7em}
	\begin{tabular}{  c  c  c  c  c  c  c  c  c  c  c  c }
		\hline\hline
		Parameter & {Real} & \multicolumn{2}{c}{Simple} & \multicolumn{2}{c}{Intermediate} & \multicolumn{2}{c}{Full (ORBIT)} 
		& \multicolumn{2}{c}{ORBIT+QPT} & \multicolumn{2}{c}{Decoherence}\\ [0.5ex]
		Learning & Model & Before & After & Before & After & Before & After & Before & After & Before & After \\ [0.5ex]
		\hline
		$\omega^{(A)}$ (MHz) & 5000 & $-$1.000 & $-$0.886 & $-$1.000 & $-$0.230 & $-$0.230 & +0.004 & +0.004 & $-$0.016 & " & " \\ [0.2ex]
		$\delta^{(A)}$ (MHz) & 210 & +1.000 & +0.702 & +1.000 & +0.281 & +1.000 & +0.400 & +0.4008 & +0.017 & " & " \\ [0.2ex]
		$\omega^{(B)}$ (MHz) & 5600 & +1.000 & +0.592 & +1.000 &  +0.013 & +0.013 & $-$0.003 & $-$0.003 & +0.006 & " & " \\ [0.2ex]
		$\delta^{(B)}$ (MHz) & 240 & +1.000 & +0.981 & +1.000 & +4.32 & +1.000 & $-$0.016 & $-$0.016 & $-$0.005 & " & " \\ [0.2ex]
		\hline
		$\varphi_0$ (MHz/V) & 159.2 & +1.592 & $-$1.634 & +1.592 & $-$0.802 & $-$0.802 & +0.123 & +0.123 & +0.246 & " & " \\ [0.2ex]
		$g$ (MHz) & 20 & -- & -- & +1.000 & $-$0.665 & $-$0.665 & +0.046 & +0.046 & $-$0.119 & " & " \\ [0.2ex]
		$T$ (mK) & 50 & -- & -- & -- & -- & +5.000 & $-$3.172 & $-$3.172 & $-$0.216 & " & " \\ [0.2ex]
		\hline
		$T^{(A)}_1$ ($\mu$s) & 27 & -- & -- & -- & -- & +4.000 & +4.439 & +4.439 & +0.021 & +0.021 & +0.738\\ [0.2ex]
		$T^{(A)}_{2*}$ ($\mu$s) & 39 & -- & -- & -- & -- & +2.000 & +1.994 & +1.994 & $-$2.353 & $-$2.353 & $-$0.020\\ [0.2ex]
		$T^{(B)}_1$ ($\mu$s) & 23 & -- & -- & -- & -- & +3.000 & +4.543 & +4.543 & +5.704 & +5.704 & +0.666\\ [0.2ex]
		$T^{(B)}_{2*}$ ($\mu$s) & 31 & -- & -- & -- & -- & +5.000 & +6.183 & +6.183 & +4.716 & +4.716 & +0.897\\ [0.2ex]
		\hline
		$p_{0 \rightarrow  0}^{(A)}$ (\%) & 97 & -- & -- & -- & -- & $-$2.00 & $-$0.84 & $-$0.84 & $-$0.11 & " & " \\ [0.2ex]
		$p_{1 \rightarrow  1}^{(A)}$ (\%) & 96 & -- & -- & -- & -- & +0.20 & +0.38 & +0.38 & +0.02 & " & " \\ [0.2ex]
		$p_{0 \rightarrow  0}^{(B)}$ (\%) & 96 & -- & -- & -- & -- & $-$2.00 & $-$0.62 & $-$0.62 & $-$0.03 & " & " \\ [0.2ex]
		$p_{1 \rightarrow  1}^{(B)}$ (\%) & 95 & -- & -- & -- & -- & +0.20 & +0.08 & +0.08 & +0.01 & " & " \\ [0.2ex]
		\hline\hline
	\end{tabular}
	\label{tab:parameters}
\end{table*}

\begin{figure}[t]
	\includegraphics[width=0.47\textwidth]{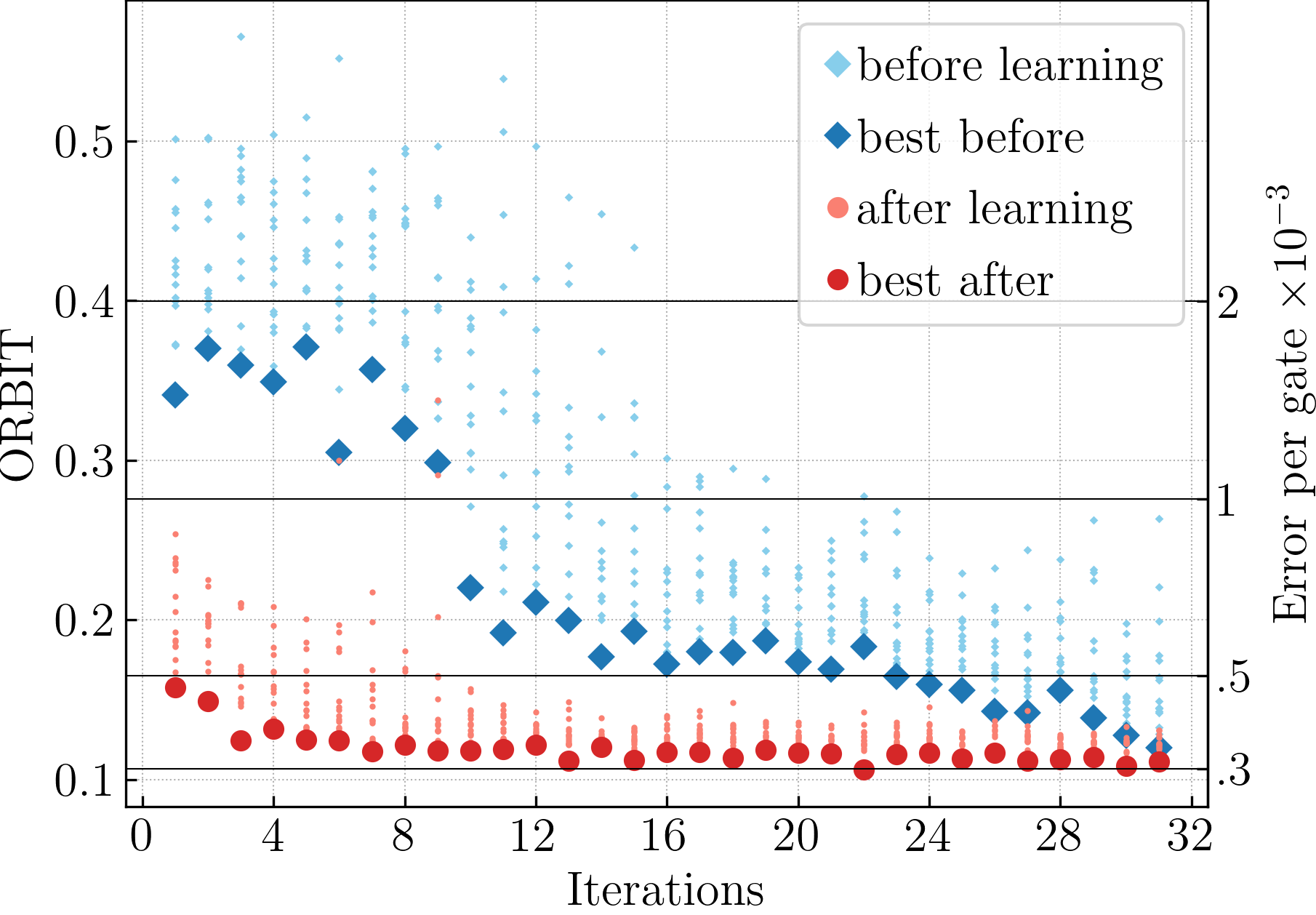}
    \caption{\cTwo\ calibration on the device for single-qubit gates of qubit $B$.
        The initial point is suggested by \cOne\ before (after) learning of the model.
        The light blue diamonds (light red circles)
        represent the values of the ORBIT goal function, Eq. (\ref{eq:ORBIT}), for varying pulse parameters $\alpha$ as
        chosen by the search algorithm. The larger blue diamonds (larger red circles) highlight the best of $25$ points generated and sampled at each iteration. 
In experimental practice, this batching helps reduce the overhead of loading pulses in AWG
programming \cite{Werninghaus2021}.
        Both calibrations achieve the same final fidelity, however the optimal gates derived from
        the learned model provide a better initial guess. 
        Assuming no SPAM errors the ORBIT value can be translated into an error per gate, indicated on the right axis. 
        This is only meant to provide a rough estimate of the performance of the gate, noting that an ORBIT value of 0.5 represents maximum error per gate, i.e. completely depolarizing channels.
    }
    \label{fig:synth-c5}
\end{figure}

\begin{figure}[t]
	\centering
    \includegraphics[width=\linewidth]{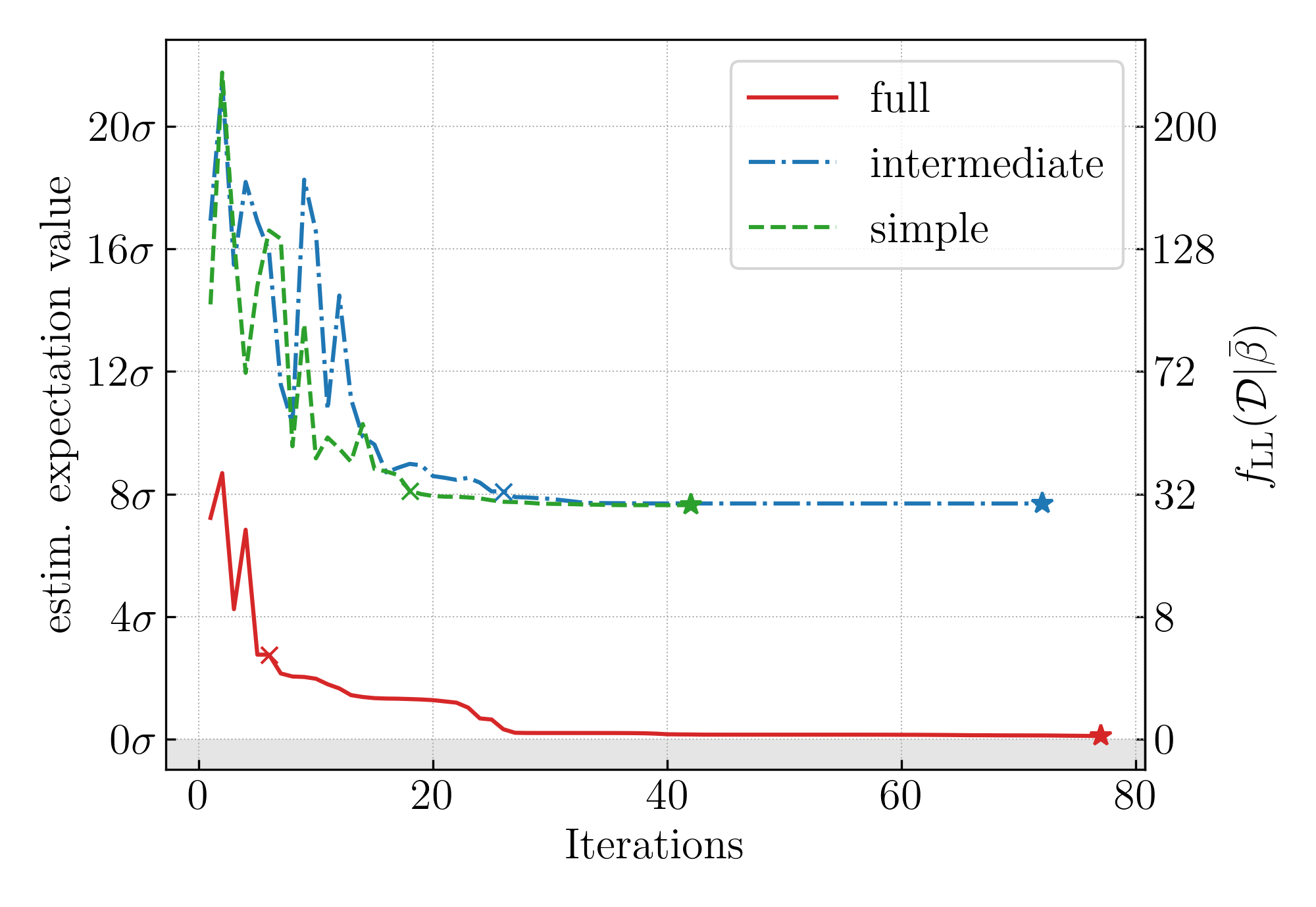}
    \caption{Progress of the \cThree\ optimization on a hierarchy of models:
    Simple model\ (green, dashed), intermediate model (blue, dot-dashed) and
    full model (red, solid), as described in the text. The Model match goal function
    $f_\mathrm{LL}(\beta)$ is defined in Eq. (\ref{eq:loglikelihood}). 
    The crosses show the switch-over from CMA-ES to L-BFGS.
    The CMA-ES algorithm evaluates a batch of points for each iteration (8, 9 and 12 for the simple,
    intermediate and full model respectively), only the best of each batch is shown.
    The L-BFGS algorithm takes on average approximately 1.2 evaluations per iteration,
    for all three models.
    The function $f_\text{LL}$ is rescaled to express the match in terms of standard deviations of
    the binomial distribution that the experimental results are drawn from. 
    The simple model is a close dispersive approximation of the intermediate model, demonstrated by
    their similar final match score.
    By including all relevant device properties the full model reaches an almost perfect match score.
    }
    \label{fig:synth-c3}
\end{figure}
\subsection{Open-loop Optimal Control, \cOne}
We assume that at the start of the \CCC\ procedure the parameters of the system are only known to a rough precision, with its qubit frequencies and anharmonicities chosen to be within a few MHz of their ``true'' values.  
In the simple model, the qubits
are \emph{uncoupled} three-level Duffing oscillators, evolution follows \emph{closed} systems
dynamics, and state preparation and measurement are assumed perfect.  The Hamiltonian is
\begin{equation}
    \begin{aligned}
    H/\hbar=&\sum_{i=A,B} \omega_i b_i^\dagger b_i - \frac{\delta_i}{2} \qty(b_i^\dagger b_i - 1) b_i^\dagger b_i \\
    &+ c_i(t) \qty(b_i+b_i^\dagger).
    \end{aligned}
\end{equation}
Assuming this model, we design pulses for single-qubit gates using \cOne.
To mitigate leakage, we choose a pulse ansatz with a Gaussian shape and a correction given by the derivative removal by adiabatic gate (DRAG) method \cite{DRAG},
\begin{equation}\label{eq:Gaussian-with-DRAG}
  \begin{aligned}
    \varepsilon(t)  &= A\,\Omega_\text{Gauss}(t)\,\cos((\omega_d+\omega_\text{off}) t + \phi_{xy})\\
            &- \frac{\eta}{\delta}\,A\,\dot{\Omega}_\text{Gauss}(t)\,
                \sin((\omega_d+\omega_\text{off}) t + \phi_{xy}).
  \end{aligned}
\end{equation}
Here, $\Omega_\text{Gauss}$ is a Gaussian envelope,
$\dot{\Omega}_\text{Gauss}(t)$ is its time derivative, $A$ is the amplitude of the drive,
$\omega_\text{off}$ is a frequency offset and the DRAG parameter $\eta$ can be adjusted to reduce leakage into the second excited state \cite{DRAG,chen-drag}.
The rotation axis can be freely chosen in the $x$-$y$ plane by changing the phase of the drive signal
$\omega_d t \rightarrow \omega_d t + \phi_{xy}$,
 implementing the unitary rotations
$R(\phi_{xy},\theta)=\exp{-i(\cos{\phi_{xy}}\sigma_{x}+\sin{\phi_{xy}}\sigma_{y})\theta}$. 
By setting $\phi_{xy}=n\frac{\pi}{2}$ with $n=0,1,2,3$ and changing $\alpha=\qty(A, \eta,
\omega_\text{off})$ we aim to realize the single qubit gate-set
\begin{equation}
\begin{aligned}
    \mathcal{G} &= \qty{X_{\pi/2}, Y_{\pi/2}, X_{-\pi/2}, Y_{-\pi/2} } 
    \label{eq:gates} \ ,
\end{aligned}
\end{equation}
for each qubit separately, eight gates in total, where $X_{\pi/2}= \qty{R(0,\pi/2)}$ and so on.
With \cOne\ we use a gradient-descent method to find the parameters $\alpha$ that minimize the mean
average gate infidelity
\begin{equation}
f_1(\alpha)=1-\frac{1}{\abs{\mathcal{G}}}\sum_{U\in\mathcal{G}} f_\text{av}(U) =1- \frac{1}{\abs{\mathcal{G}}}\sum_{U\in\mathcal{G}} \frac{\chi_{0,0} d + 1}{d+1},
\end{equation}
where, $\chi_{0,0}$ is the $(0,0)$-th element of the Chi matrix representation of the gate error 
$U^\dagger \circ \widetilde{U}(\alpha)$ between the ideal gate $U$ and the implemented gate
$\widetilde{U}(\alpha)$ \cite{emerson-avfid}.
\begin{figure*}[t]
	\centering
	\includegraphics[width=\linewidth]{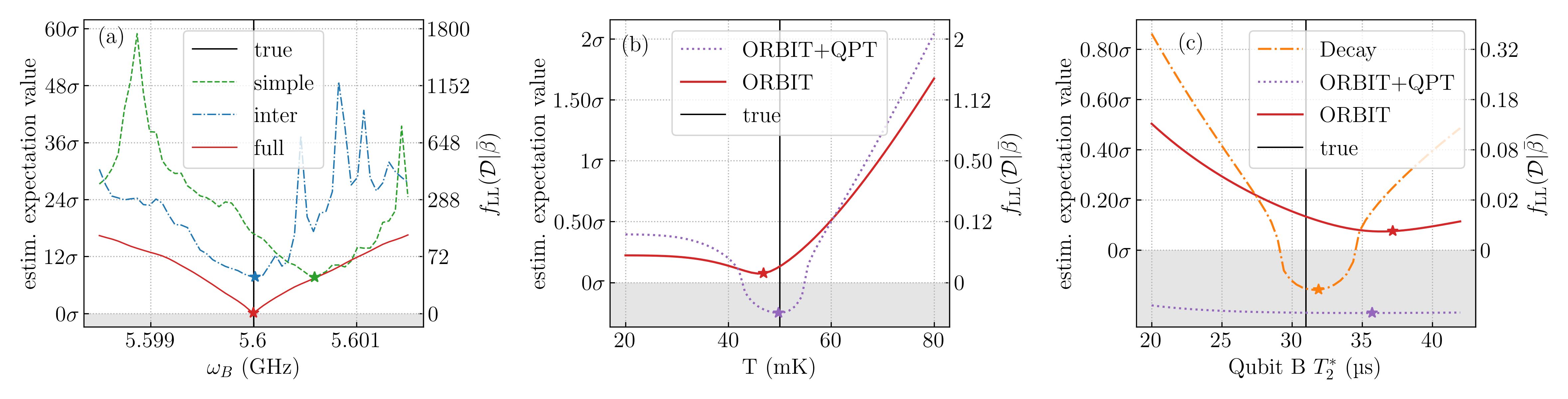}  
	\phantomsubfloat{\label{fig:sens-freq}}
	\phantomsubfloat{\label{fig:sens-temp}}
	\phantomsubfloat{\label{fig:sens-t2}}
	\caption{
	Sensitivity analysis of selected parameters for different models and data sets.
	In (a) we sweep the qubit frequency $\omega_B$ and evaluate the goal function 
	$f_\mathrm{LL}$ on the learning data at each value. The star represents the optimal value
	returned by \cThree. Intermediate (blue, dot-dashed) and Full (red, solid) models, show the same
	frequency value, while the value of the Simple model (green, dashed) is dispersively shifted, as expected.
	In (b) and (c) we perform the same sweep for the chip temperature and $T_2^*$ of qubit $B$ respectively, 
	evaluated on the full model for different learning data-sets.
	The ORBIT data is the same used in the full model in (a).
	Introducing the Quantum Process Tomography (QPT) data (purple, dotted), allows a more precise definition of the temperature.
	To determine $T_2^*$ we use relaxation and dephasing data (orange, dot-dashed). 
	Match values below 0 can occur because of noisy data, finite sampling and deviation from the assumption of
	Gaussian distribution of the data.
	More sensitivity plots are shown in the Supplemental Material \cite{supp}.
	}
\end{figure*}
We optimize Gaussian pulses with a gate length of $t_g=7~\rm{ns}$, for both qubits, using the gradient-based L-BFGS algorithm \cite{L-BFGS}.
The obtained optimal pulses yield a mean infidelity of $f_1(\alpha)= \EE{6.6}{-4}$ 
and $f_1(\alpha)= \EE{4.9}{-4}$ on the simple model for qubit $A$ and qubit $B$ respectively, realistic values for fast gates using this simple parametrization.
Next, we compare the performance of these pulses on the black-box device, where the gates instead yield a mean infidelity of $\EE{2.4}{-3}$ for qubit $A$ and 
$\EE{1.5}{-3}$ for qubit $B$.
In fact, performing an experimentally realistic randomized benchmarking (RB) \cite{Knill09, Chow2009, Magesan2011, Magesan2012} measurement on the device yields an error per gate of
$\EE{2.3}{-3}$ and $\EE{1.3}{-3}$ comparable with the theoretical average infidelity.
The degradation of performance from optimal control simulation ($\approx 10^{-4}$) to experiment ($\approx 10^{-3}$) shows a clear mismatch 
between the device and the simple model.

\subsection{Calibration, \cTwo}
The next step is to calibrate the pulses derived by \cOne\ and improve their performance on the
device. 
We use \cTwo\ and employ a closed-loop, model-free, gradient-free optimization
algorithm on an experimentally accessible figure of merit $f_2$. Since the goal is to evaluate a gate-set, we choose $f_2$ to be the ORBIT
\cite{ORBIT} (single-length RB) function
\begin{equation}\label{eq:ORBIT}
    f_2(\alpha)=f_\text{ORBIT}(\alpha) = \frac{1}{N}\sum_{k=1}^N \qty[1 - m_k(\alpha)],
\end{equation}
averaging over $N$ sequences. 
The survival probability, $m_k = \Pi_0\qty(S_k(\alpha)\rho_\text{init} S_k^\dagger(\alpha))$, is the probability to
measure the state $\ket{0}$ (see Eq. \ref{eq:readout})
after applying random sequences 
\begin{equation}
    S_k\qty(\alpha) :=\qty{\prod_j^{L-1} C_{k,j}}C_\text{inv}
\end{equation}  
composed of $L$ Clifford gates \cite{ORBIT}, to the initial thermal state $\rho_\text{init}$.
The $C_{k,j}$ are the random gates sampled from the
Clifford group $C$ (for a single qubit $\abs{C}=24$), and $C_\text{inv}$ is chosen so that $
S_k\equiv \mathcal{I} $ in the ideal case.
We use the atomic operations  $\mathcal{G}$ from Eq. (\ref{eq:gates}) to construct the set of
Clifford gates, e.g. $C_6=X_{-\pi/2}\circ Y_{-\pi/2}\circ X_{\pi/2}$, and from them construct $N=25$ RB sequences of length  $L=100$.
The survival probabilities $m_k$ are estimated by performing $s=1000$ projective measurements and averaging.

To minimize $f_2$, we employ the CMA-ES \cite{CMA-ES} algorithm, a gradient-free search that samples
the loss function in batches, and is fairly robust to local minima and noise \cite{Sinanovic-Thesis}.
See \cite{Werninghaus2021} for an experimental demonstration.
The optimal pulse parameters from \cOne\ are used as the starting point of the optimization, and the
parametrization is kept as in Eq. (\ref{eq:Gaussian-with-DRAG}).
We perform the calibration for each qubit independently, with similar results. See Fig. \ref{fig:synth-c5} for the ORBIT calibration data of qubit $B$. The initial point suggested by \cOne\ 
has an ORBIT infidelity of $0.50$ and is improved by the optimization to $0.12$.
To account for SPAM errors, we perform a full RB measurement and estimate the infidelity
of the gates  before and after as $\EE{1.3}{-3}$ and $\EE{3.4}{-4}$ respectively. Qubit $A$ shows a
similar improvement of RB estimated error from $\EE{2.3}{-3}$ to $\EE{7.5}{-4}$.

For the purpose of learning we define the data-set $\mathcal{D}:=\qty{S_k(\alpha_j)\rightarrow m_{j,k}}$, the collection of
the experiments conducted during the \cTwo\ calibration, consisting of pulse parameters
$\alpha_j$ and gate sequences $S_k(\alpha_j)$, and the corresponding measurement outcomes $m_{j,k}$.

\subsection{Characterization, \cThree}

\begin{figure*}[t]
	\includegraphics[width=0.95\linewidth]{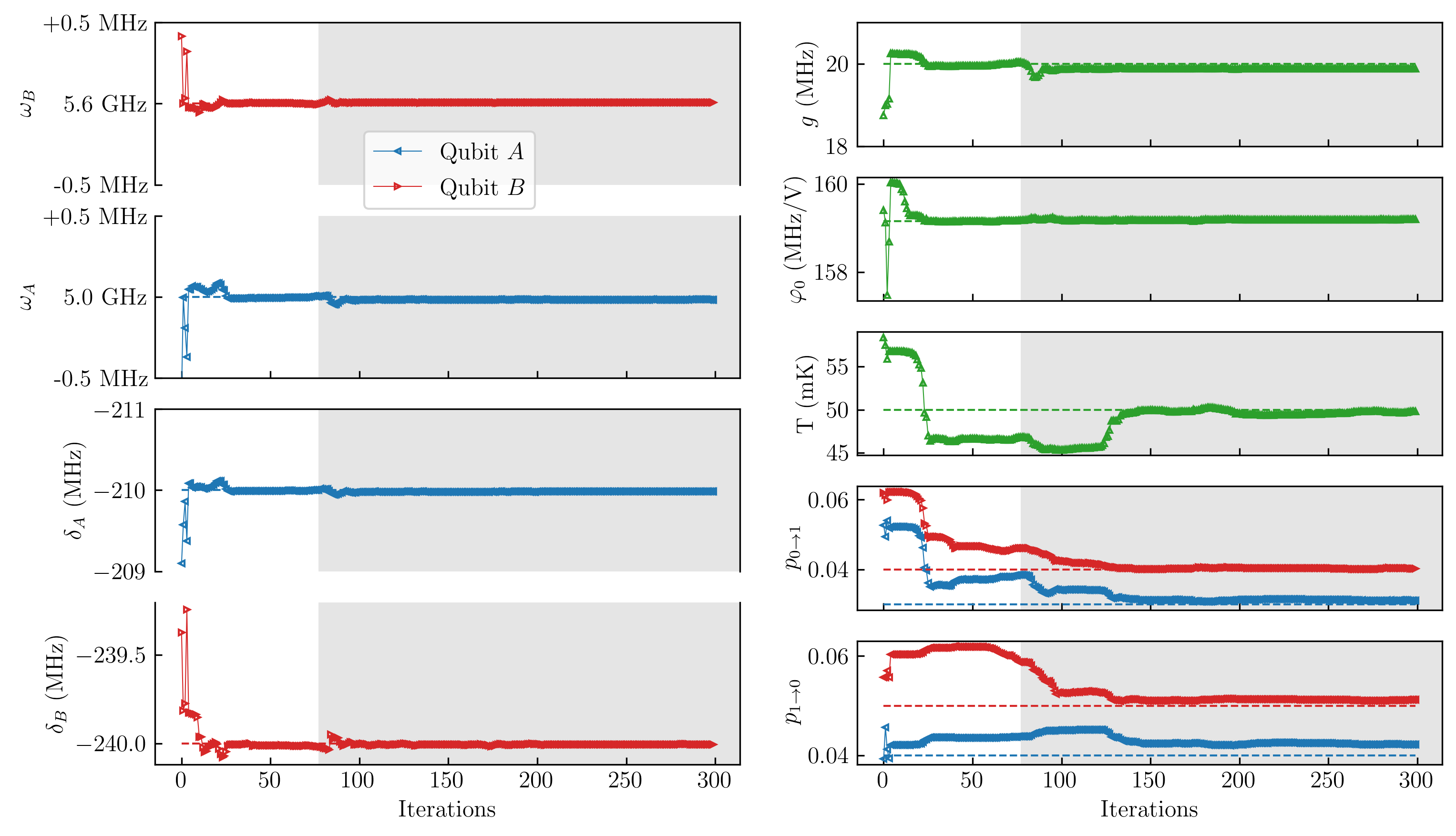}
	\phantomsubfloat{{\label{fig:synth-c3-par}}}
	\caption{
	    \cThree\ learning of the two-qubit model parameters. Blue, left (red, right) triangles indicate
		qubit $A$ ($B$) parameters, respectively, while shared properties are shown with green
		upwards triangles. The true values of the ``real'' model are indicated as dashed lines.
		Learning begins using just ORBIT data (left white section) that fixes qubit frequencies,
		anharmonicities, coupling and line transfer function to their true values. 
		Then tomography data from a two qubit experiment is added (right gray section), 
		which allows better identification the chip temperature $T$ and the misclassification constants
		$p_{0\rightarrow1}$, and $p_{1\rightarrow0}$.
	}
\end{figure*}

In \cThree, we use the data-set $\mathcal{D}$ obtained during ORBIT calibration to improve the model of the system.
For each measurement result $m_{j,k}$ we compute the equivalent simulation result 
$\widetilde{m}_{j,k}(\beta)$ by calculating the dynamics of the sequence $S_k(\alpha_j)$
given a set of  model parameter values $\beta=(\omega_i, \delta_i, ...)$.
Since simulating the whole data-set is computationally costly, for the purpose of model learning we make a selection 
of eight pulse parameter sets $j$ per qubit from the full data-set. 
Each parameter set includes $k=1,..,25$ sequences, meaning that we learn from a total of $N=400$ measurement results, relabeled as $m_n$. We then construct a goal function 
\begin{equation}\label{eq:loglikelihood}
    f_3(\beta) = f_\text{LL}\qty(\mathcal{D} \vert \beta) =
        \frac{1}{2N}\sum_{n=1}^{N}\qty[\qty(\frac{m_n-\widetilde{m}_n}{\widetilde{\sigma}_n})^2-1]
\end{equation}
that captures how well the model prediction $\widetilde{m}_{n}$, with standard deviation $\widetilde{\sigma}_n$, agrees with the recorded values $m_{n}$.
Because of the finite number of measurements, the averaged $m_n$ are noisy estimates of the population, with a mean $\mu_n$ and standard deviation $\sigma_n$.
Thus, they cannot be matched perfectly even when all model parameters are exact. 
However, we can determine the expectation value of the goal function $f_\text{LL}$ in the scenario where   
all $\widetilde{m}_n$ are exactly a given number of standard deviations away from the underlying true value $\mu_n$. 
A detailed mathematical discussion is presented in Sec. \ref{sec:f_LL}.
To provide a more intuitive measure, we express the match $f_\text{LL}$ in terms of the number of standard deviations that would result in the same score. 

To minimize $f_3(\beta)$, we use a combination of two algorithms:
Gradient-free (CMA-ES) to avoid local minima and gradient-based (L-BFGS) to converge quickly once
the right minimum has been identified.
Fig. \ref{fig:synth-c3} shows the convergence of the \cThree\ optimization for different models. 
The simple model is not able to reproduce the device's results, as the optimization ends at
approximately 8 standard deviations away. 
This demonstrates that the experiment on the device includes behavior not captured by the simple model. 
Spectator effects might be significant even when performing only single qubit experiments, 
making the completely uncoupled model insufficient.
Another source of this inconsistency might be SPAM errors not accounted for in the model,
that might play a large role in actual measurement results.
The parameter values resulting from this \cThree\ process and all following ones are shown in Table \ref{tab:parameters}. 

Going forward an informed decision has to be made about how to enhance the model. Since the true values of the parameters are not known in an experimental setting, we require a tool to determine 
the precision to which they are learned. 
We estimate the sensitivity to changes of model parameters around the optimal values $\beta'$ by
performing one-dimensional scans and observing the degradation in model match score,
$f_\text{LL}(\mathcal{D} \vert \beta'+\delta\beta)$. Fig. \ref{fig:sens-freq} shows that sweeping the value of frequency of qubit $B$ produces a highly irregular landscape of the match score $f_\text{LL}$.

The simple model is then extended by adding the static coupling $g$ of unknown exact value, resulting in the intermediate model. When repeating \cThree, we initialize model parameters from the initial, rough values. We do not carry over the learned parameters
from the simple model to the intermediate model because, by introducing
a coupling, we expect slightly shifted frequencies compared to the bare frequencies of the uncoupled qubits.
Nonetheless, convergence of the match score shows no improvement from the simple model, still only reaching within approximately 8 standard deviations from experiment results (Fig. \ref{fig:synth-c3}) and resulting in a similar sensitivity landscape in Fig. \ref{fig:sens-freq}.
This suggests that the simple model is a close dispersive approximation of the intermediate model.
Indeed, we observe a dispersive shift \cite{Blais2004} of $593~\rm{KHz}$, consistent with the expected 
$g^2/(\omega_B-\omega_A)\simeq 666~\rm{KHz}$, given the coupling of $g \simeq 20~\rm{MHz}$ and the frequency difference
$\omega_B-\omega_A\simeq 600~\rm{MHz}$. 

Finally, model complexity is increased by adding three relevant features: Markovian noise simulated
by Lindblad master equation, initialization errors due to finite operating temperature and
measurement errors in the form of misclassification.
The system model is now of the same structure as the ``real'' model of the device. Starting from the best intermediate
model parameters, the \cThree\ procedure converges satisfactorily,
approaching the 0 standard deviations mark (Fig. \ref{fig:synth-c3})

In Fig. \ref{fig:synth-c3-par} we show the value of each parameter of the full model during
optimization, as we introduce different learning data (in the next sections), and compare with 
their true value (dashed lines). 
By learning the model parameters with the ORBIT data (white left section of each plot) the model frequencies $\omega_{A/B}$, anharmonicities $\delta_{A/B}$, coupling $g$ and line transfer function $\varphi_0$ converge to their true value.
The temperature and misclassification parameters are not recovered, and we believe this is due
to an extra degree of freedom that is not resolved by the experiments we have performed, 
as the effects of misclassification, Eq. (\ref{eq:readout}), and initial thermal distribution, Eq. (\ref{eq:thermal}),
are similar and can be partially exchanged.
Dephasing and relaxation times (not shown) are also not recovered. Indeed, in Fig. \ref{fig:sens-t2} we show that the sensitivity of the data to dephasing time $T_2^*$ of qubit $B$ is minimal. 
RB sequences perform an effective random dynamical decoupling
\cite{wallman2014randomized}, providing a possible explanation to this result.
\begin{figure*}[t]
	\includegraphics[width=\linewidth]{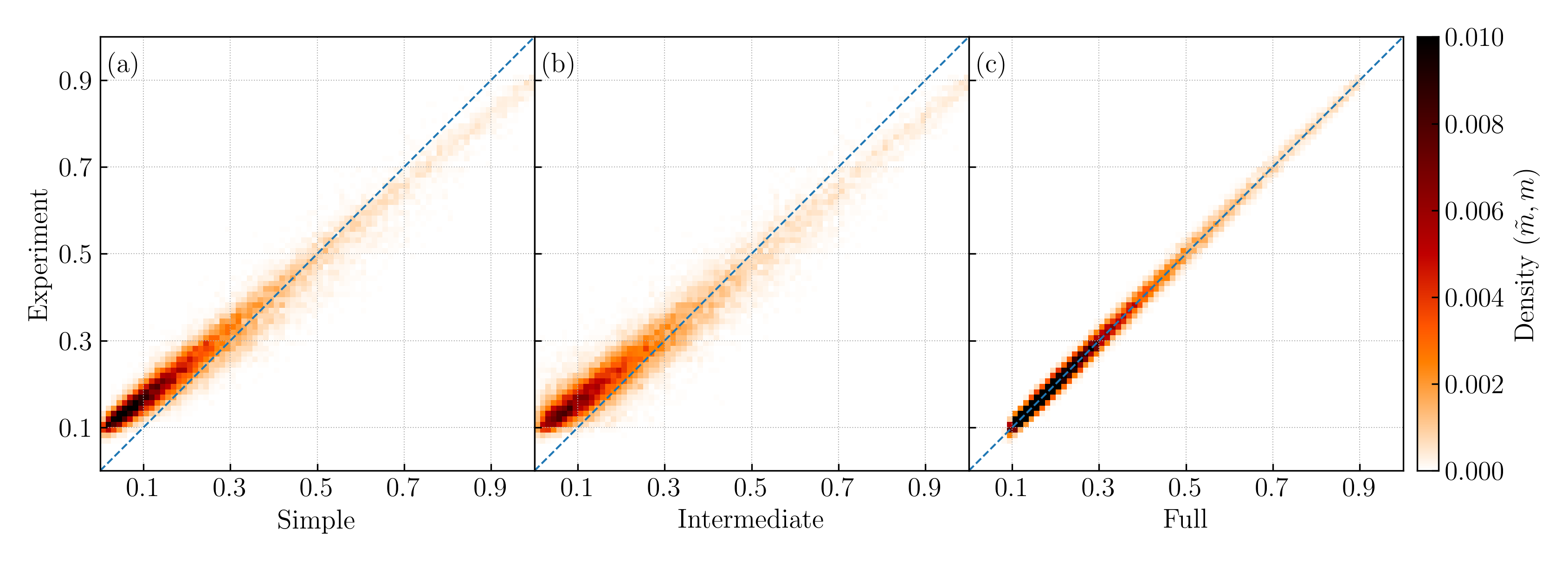}
	\phantomsubfloat{\label{fig:hist-c3-1}}
	\phantomsubfloat{\label{fig:hist-c3-2}}
	\phantomsubfloat{\label{fig:hist-c3-3}}
	\caption{
		(a)-(c): Correlation between simulation and experiment expressed as a density of points
		$(\widetilde{m}_k, m_k)$ for the Simple, Intermediate and Full model, respectively. The
		data-set is the so-called validation set: The data points $k$ that were not used in
		optimization of the model parameters. The simple and intermediate models show poor
		correlation as the simulation predicts a wide distribution of measurement outcomes for each
		recorded value. They also exhibit a tilt that can be attributed to SPAM errors not considered
		in the models. Only the full model produces a consistently high density distribution centered 
		around the diagonal, with minimal spread due to the noisy data.
	}
	\label{fig:hist-c3}
\end{figure*}

\subsection{Validation of the learned model}
After model matching on a subset of the data in the \cThree\ step, we now evaluate the predictive power of the learned
models by computing the score on the rest of the data set (this is also known as a validation set in machine learning). 
This verifies that the selected subset captures all relevant behavior and alleviates the danger of overfitting.

\Cref{fig:hist-c3-1,fig:hist-c3-2,fig:hist-c3-3} depict the correlation between calibration data points $m_{j,k}$ and
their model-based reconstructions $\widetilde{m}_{j,k}$. We evaluate the goal function $f_\text{LL}(\beta)$ over the
validation set for the Simple, Intermediate and Full models and obtain values of 36.5 ($\approx
8.4 \sigma$), 42.0 ($\approx 9.2 \sigma$) and 0.028 ($\approx 0.2 \sigma$) respectively.
The conclusion is that, even though some parameters were not recovered by \cThree, the learned full model is indeed a Good Model, as it reproduces the behavior of the system on \emph{all
previously recorded data points} to satisfying accuracy.
This does not prevent additional measurement data to expose new behavior in the system: 
The notion of the Good Model is always tied to the underlying data-set.

Furthermore, we now repeat the \cOne\ procedure on the Good Model 
(yielding average gate infidelities of $\EE{6.3}{-4}$ and $\EE{1.1}{-3}$ for qubit $A$ and $B$ respectively)
and show that the resulting pulses give a near optimal performance on the actual device and allow for 
faster \cTwo\ convergence, as seen in Fig. \ref{fig:synth-c5}. One would expect the pulses derived from the Good Model to be exactly optimal on the actual device. 
Even though it is not the case here, this is not because of an inaccurate model, but rather because of a 
disparity between the figures of merit used in \cOne\ (average infidelity) and \cTwo\ (single-qubit ORBIT).
Average fidelity captures effects of the whole system, including, in this case, an effective ZZ-coupling between 
the two qubits caused by a slight repulsion of the $\ket{02}$ and $\ket{11}$ states,
that are $300~\rm{MHz}$ apart. 
Minimizing a single-qubit ORBIT infidelity does not adjust for this effect, as we can verify by evaluating both RB (which captures only one qubit at a time) and average infidelity before and after calibration.
Indeed, the average infidelity of qubit $B$ is $\EE{1.2}{-3}$ (compatible with the performance of $\EE{1.1}{-3}$ on the 
Good Model) but the error per gate is estimated by RB as $\EE{4.1}{-4}$.
After the calibration the RB estimate is improved to $\EE{2.9}{-4}$ but the average infidelity is
worsened to $\EE{1.9}{-3}$.
Performing simultaneous RB could resolve this issue.

\subsection{Entangling gate}\label{two-qubit-calibration}
We further investigate the Good Model which was determined using only single-qubit calibration data 
by deriving a two-qubit cross-resonance (CR) gate \cite{rigetti2010, chow2011} with \cOne. 
Both qubits are driven simultaneously at $\omega_B$, the
resonant frequency of qubit $B$, to accumulate a phase $\pm \pi/2$ conditioned on the state of qubit 
$A$ \cite{Sheldon2016}. Both drives are parameterized by flattop Gaussians. 
The resulting CR pulse has a gate infidelity of $f_\text{av}=\EE{3.8}{-3}$.
When evaluated on the ``real'' model the gate has an infidelity of $f_\text{av}=\EE{4.3}{-3}$, again
showing that the learned model predicts device behavior to high accuracy. Notably, the model learned 
using only single-qubit data was sufficient to accurately predict the performance of the two-qubit gate
on the device. We suspect this to be caused by exchange interactions due to coupling and finite temperature: Even when performing only single-qubit gates, the finite temperature causes a partial excitation of higher states, which are then exchanged with the other qubit via the coupling and thus visible in the ORBIT data.

The performance of the gate on the device is verified with Quantum Process Tomography (QPT): We
apply the CR gate preceded and followed by single-qubit gates to prepare and measure in the basis
states, e.g. $S = \qty(X_{\pi/2} \otimes Y_{\pi/2}) \circ \text{CR} \circ \qty(X_{-\pi/2}\otimes Y_{\pi/2})$ \cite{Alexander2020}, and again collect these measurements into our data-set. 
We believe that the entangling gate lifts the degree of freedom between misclassification and initial thermal distribution discussed before, 
hence we now perform another \cThree\ optimization, using the QPT data ($256$ sequences) and one ORBIT parameter per qubit ($2\times25$ sequences) as the learning data.
Parameter convergence is shown in the gray areas of Fig. \ref{fig:synth-c3-par}, where temperature 
and confusion matrix values are adjusted closer to the true values.

Fig. \ref{fig:sens-temp} substantiates the claim that the entangling gate data allows for a more precise learning of the chip temperature, 
exhibiting a narrower valley at the true value. However, we are still not able to learn the $T_1$ and $T_2^*$ parameters, 
since the sequences in QPT are too short to be sensitive.

\subsection{Relaxation and dephasing}
To demonstrate how a specialized measurement is formulated within \CCC\, we determine the values of $T_1$ and $T_2^*$, using simple
established sequences that are known to be sensitive to these parameters. The decay lifetime $T_1$ is
determined by preparing the excited state of the qubit, followed by increasing wait times and then
measuring the ground state population. We write the sequence as 
\begin{equation}\label{eq:decay}
    S_{T1}^{(n)} = X_{\pi/2} \circ X_{\pi/2} \circ \mathcal{I}^n
\end{equation} 
where $X_{\pi/2}$ is our previously optimized $\pi/2$ gate and $\mathcal{I}^n$ signifies $n$
repetitions of the identity gate $\mathcal{I}$.
Similarly 
\begin{equation}\label{eq:ramsey}
    S_{T2*}^{(n)} = X_{\pi/2} \circ \mathcal{I}^{n/2} \circ X_{\pi/2} \circ X_{\pi/2} \circ \mathcal{I}^{n/2} \circ X_{-\pi/2}
\end{equation}
defines a Ramsey echo sequence, used to measure the dephasing time $T_2^*$.
We take 51 logarithmically spaced values of $n$ between 100 and 10000 to capture the full decay curves.

\begin{figure}[t]
	\centering
	\includegraphics[width=0.95\linewidth]{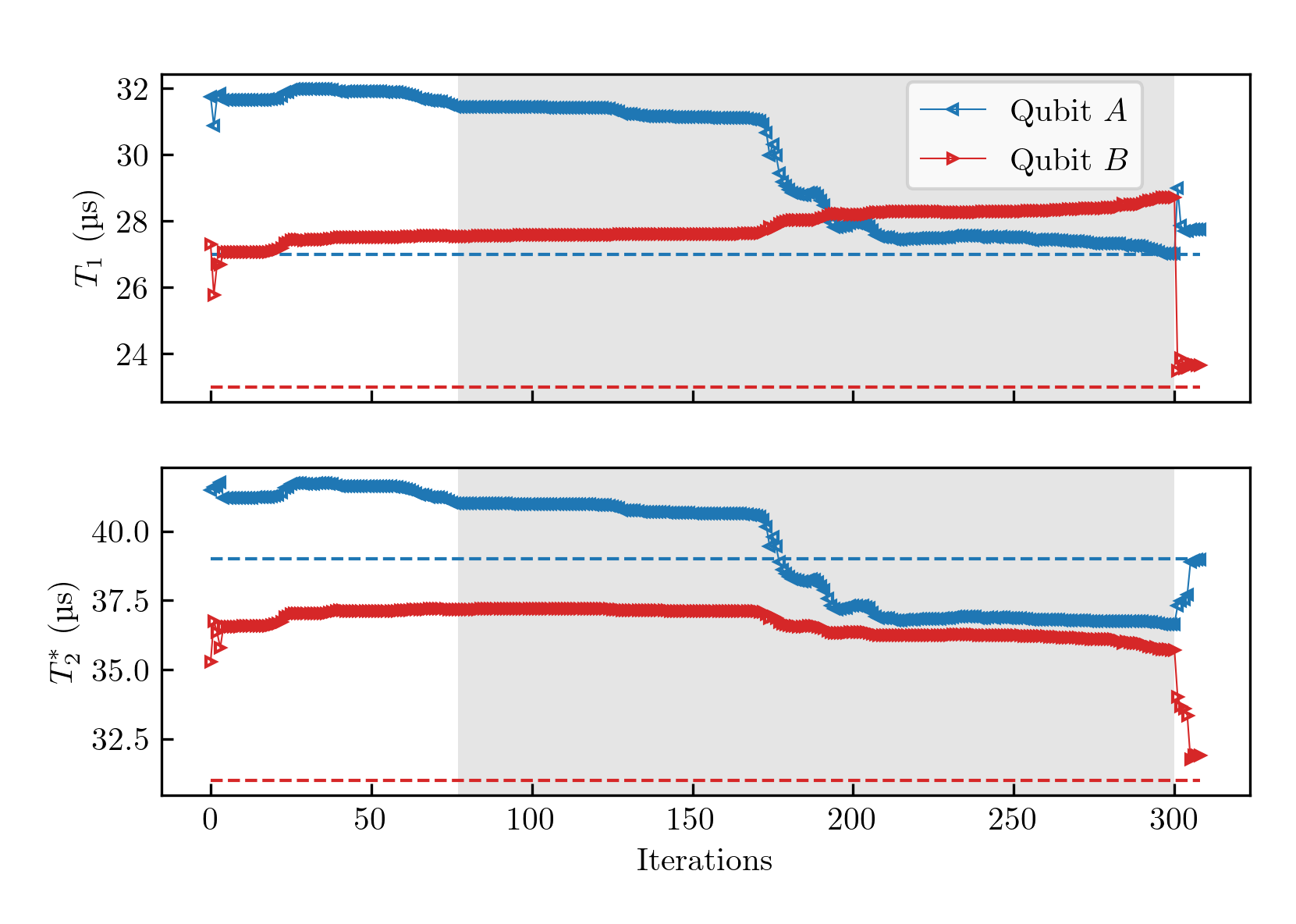}
	\caption{\cThree\ learning of the relaxation ($T_1$) and dephasing ($T_2^*$) parameters. Blue, left (red, right) triangles indicate
		qubit $A$ ($B$) parameters, respectively. The true values of the ``real'' model are indicated as dashed lines.
		Background sections represent different learning datasets: just ORBIT data (left white section), a mix of ORBIT and QPT data (center gray section), decay and Ramsey data (right white section).
		The decay times are correctly identified only when specific data, sensitive to the decoherence effects, is used for learning, at which point they quickly converge to the real value.
	}
	\label{fig:t1_t2}
\end{figure}

Using this data-set we perform another \cThree\ optimization, freezing all model parameters
learned until now and varying only the values of $T_1$ and $T_2^*$.
By doing so we manage to determine their values to within $1\mu$s of the true values (Fig. \ref{fig:t1_t2}).
This procedure is the \CCC\ equivalent of a common exponential decay fit to the data.
However, with \CCC\ one does not require prior knowledge on the expected structure of the 
experimental results, i.e. an exponential decay. Hence, when matching the data \CCC\ also 
accounts for SPAM errors without the need to adjust the fitting function.

Fig. \ref{fig:sens-t2} shows the sensitivity of $f_\text{LL}$ to the value of $T_2^*$ of qubit $B$.
The new data shows a clear improvement in the accuracy of the value obtained and the minimum is 
better defined. For increased sensitivity one would require more decay data to learn from.

\subsection{Sources of error}\label{example-error-budget}

The Good Model allows us to break down which of the model properties are preventing higher gate
fidelities. To this end, we investigate the Good Model for components limiting the performance of 
the CR gate by idealizing aspects of the model. 

We investigate whether the Gaussian ansatz is limiting gate fidelities by further refining the optimal
pulses using a piece-wise constant optimization with one pixel per AWG sample (as is done in \cite{Werninghaus2021}).
Average infidelity improves only marginally from $f_\text{av}^\text{DRAG} = \EE{3.8}{-3}$ to
$f_\text{av}^\text{PWC}= \EE{3.6}{-3}$, suggesting other factors are limiting fidelities.

To find out if performance is limited by decoherence effects, we re-optimize the CR gate while
disabling Lindbladian dynamics. By open-loop optimization in this idealized coherent setting the error is decreased
from $\EE{3.8}{-3}$ to $\EE{1.3}{-5}$. 
Thus, the $100~\rm{ns}$ CR gate considered here is coherence limited, as is the case in most experimental implementations
\cite{Sheldon2016, Corcoles2013}, making improvements in gate time essential \cite{Kirchhoff2018}.

\section{\CCC\ in-depth}\label{sec:CCC-in-detail}

Following is a detailed description of the \CCC\ tool-set, its modeling capabilities and a general formulation of the optimization problems discussed in the previous section.

\subsection{Experiment modeling}\label{sec:Simulator-core}

To combine control and characterization, \CCC\ provides
a detailed simulation that endeavors to encompass all relevant practical considerations of the experiment such as signal processing,
SPAM errors, control transfer functions and Markovian noise.
The simulator is used as the basis of the open-loop
optimal control optimization (\cOne) and the model parameter optimization (\cThree). In both cases it is desirable
to use gradient-based optimization algorithms \cite{Machnes2018,wilhelm2020introduction}. However, it is
extremely cumbersome to manually derive the full analytical gradients of the quantum dynamics, especially when it includes the properties described above. Instead, \CCC\ uses a numerics
framework \cite{tensorflow2015-whitepaper} which allows for automatic differentiation \cite{kedem1980automatic}, making the tool-set more flexible and easily extendable.
A similar approach is also used by \cite{leung2017speedup,ball2020software} for control and characterization.

\subsubsection{Signal processing}

The simulation allows for the specification of control signals $\varepsilon(t)$ as either analytical
functions or as direct, piecewise constant AWG parameterization. Analytic parametrizations are sampled at the resolution of the waveform generator producing the envelope signal
$\varepsilon_{i}=\varepsilon(t_i)$, representing voltages being applied to the control line, where the $\{t_i\}$ are the
AWG sample times. The resulting signal will exhibit a \emph{rise time} $\tau$, due to the finite bandwidth of the control electronics. We model this by applying a convolution
\begin{equation}
\begin{aligned}
\widetilde{\varepsilon}(t)&=
\int\limits_{t_0}^{t_f}\text{interp}\qty(\qty{\varepsilon_{i}})(t)G\qty{t_f-t}\dd{t}
\end{aligned}
\label{eq:convolution}
\end{equation}
with
\begin{equation}
\begin{aligned}
G(t)&=\frac{1}{N}\exp{-\frac{\qty(t-\tau/2)^2}{8\tau^2}} \ ,
\end{aligned}
\end{equation}
modeling a Gaussian filter, and
\begin{equation}
\text{interp}\qty(\qty{\varepsilon_{i}})(t) = \qty{\varepsilon_{i} \,\vert\, t_{i}\leq t < t_{i+1} }
\end{equation}
interpolating the sampled signal to higher resolution for simulation. An I/Q-Mixer combines this envelope with a local oscillator signal of frequency $\omega_\text{lo}$ to
\begin{equation}
u(t) = I(t)\cos(\omega_\text{lo} t)-Q(t)\sin(\omega_\text{lo} t)
\end{equation}
where the in-phase and quadrature components 
\begin{equation}
\begin{aligned}
I(t)&=\widetilde{\varepsilon}(t)\cos(\phi_{xy} - \omega_\text{off}t)\\
Q(t)&=\widetilde{\varepsilon}(t)\sin(\phi_{xy} - \omega_\text{off}t) 
\end{aligned}
\end{equation}
are assigned by a control parameter $\phi_{xy}$, and modulated to introduce a frequency offset $\omega_\text{off}$ on the drive. 
As noted in \cite{mckay2017efficient}, in practice there will be
additional errors during the mixing, which are not currently modeled.
In transmitting this signal to the experiment, various distortions can occur, modeled by a response
function $\varphi$, which also converts the field from line voltage to an amplitude
$c(t)=\varphi\qty[u(t)]$.

\subsubsection{Time evolution}
The system Hamiltonian is
\begin{equation}\label{eq:ctrl-ham}
H(t) = H_0 + \sum_k c_k(t) H_k \ ,
\end{equation}
with a  drift $H_0$ and optional control Hamiltonians $H_k$. 
The dynamics of the system are described by the time-ordered propagator 
\begin{equation}\label{eq:propagation}
U(t) = \mathcal{T}\exp{-\frac{\text{i}}{\hbar}\int\limits_{t_0}^t H\qty(t') \dd t'} ~ ,
\end{equation}
given by solving the time-dependent Schr\"odinger equation, and approximated numerically by $U(t) \simeq \prod_{i=N}^0 U_i $. Here, $U_i = \exp{-\frac{\text{i}}{\hbar}H(t_i)\Delta t}$, and
the total time is divided into $N$ slices of length $\Delta t$ that are fine enough so that the Hamiltonian can be considered constant in the interval.

In application, we will rarely perform a single gate or pulse in isolation. Experiments such as
randomized benchmarking or the various flavors of tomography involve long pulse sequences, that are
inefficient to simulate as a whole. Instead, the \CCC\ simulator computes each propagator $G$
of a defined gate-set $\mathcal{G}$ individually and compiles these matrix representations into sequences.
This avoids the need to solve the equations of motions multiple times for the same exact pulses.
As the propagators are calculated in the dressed laboratory frame (as opposed to the single-particle rotating frame),
consecutive gates need to be adjusted to realign with the rotating frame of the drive signal, by applying
a $Z$ rotation with an angle of $(\omega_\text{lo}+\omega_\text{off})t_g$ \cite{mckay2017efficient}.

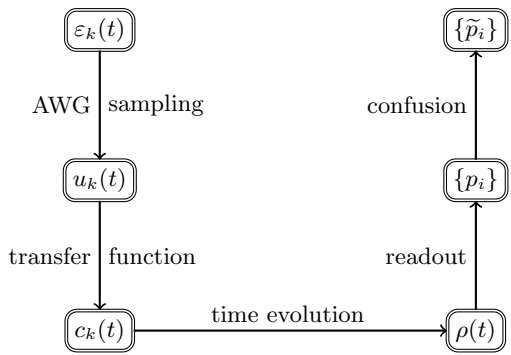
\begin{figure}[t]
	\begin{tikzpicture}
	\path (0,2) node[draw,double,rounded corners](x1) {$\varepsilon_{k}(t)$}
	-- (0,0) node[draw,double,rounded corners](x2) {$u_k(t)$} node[left, midway]{AWG}
	node[right, midway]{sampling}
	-- (0,-2) node[draw,double,rounded corners](x3) {$c_k(t)$} node[right, midway]{function} node[left, midway]{transfer}
	-- (5 ,-2) node[draw,double,rounded corners](x4) {$\rho(t)$} node[above, midway]{time evolution}
	-- (5,0) node[draw,double,rounded corners](x5) {$\{p_i\}$} node[left, midway]{readout}
	-- (5,2) node[draw,double,rounded corners](x6) {$\{\widetilde{p}_i\}$} node[left, midway]{confusion};
	\draw[->, thick] (x1) -- (x2);
	\draw[->, thick] (x2) -- (x3);
	\draw[->, thick] (x3) -- (x4);
	\draw[->, thick] (x4) -- (x5);
	\draw[->, thick] (x5) -- (x6);
	\end{tikzpicture}
	\caption{The process of simulating experimental procedure for signal processing and readout. 
		The $k$-th control function is specified by some function $\varepsilon_k(t)$ and specifies the line voltage $u_k(t)$
		by an arbitrary waveform generator (AWG) with limited bandwidth.
		Electrical properties of the setup, such as impedances, are expressed as a line transfer
		function $\varphi$, resulting in a control field $c_k(t)=\varphi\qty[u_k(t)]$, as in Eq.
		(\ref{eq:ctrl-ham}). After solving the equation of motion for the system, readout and
		misclassification are modeled by applying rescaling and transformations to the simulated
		populations $p_i=\abs{\rho_{ii}}^2$, according to Eq. (\ref{eq:confusion}).
	}
	\label{fig:flow-exp-to-sim}
\end{figure}

To include open-system effects, we apply the equivalent procedure to obtain the process matrix
\begin{equation}\label{eq:propagation-lind}
\mathcal{E}(t) = \mathcal{T}\exp{\int\limits_{t_0}^t \mathcal{L}\qty(t') \dd t'}
\end{equation}
by solving the master equation in Lindblad form,
\begin{equation}
\dot{\rho} = \mathcal{L}(\rho) = -i [H, \rho]  + \sum_{j} L_{j} \rho L_{j}^\dagger- \frac{1}{2}
\qty{L_{j} L_{j}^\dagger, \rho},
\end{equation}
where $H$ is the Hamiltonian from Eq. (\ref{eq:ctrl-ham}), the $L_{j}$s are Lindblad operators, and $\mathcal{L}$ 
is the generator in superoperator form \cite{lidar2019lecture}. The evolution of a state is obtained by applying the propagator as
$\rho_{f}=U(t_g)\rho_i U^\dagger(t_g)$ for coherent evolution or $\rho_f= \mathcal{E}(t_g)\qty[\rho_i]$ for incoherent evolution.

\subsubsection{Initialization and readout}

Given the temperature $T$ of the device, the system is initialized in a mixed state
\begin{equation}
\rho_\text{init}=\sum_k\dfrac{1}{Z}\ketbra{\phi_k}{\phi_k}\exp{-E_k / k_B T}
\end{equation}
where $\qty{\ket{\phi_k}}$ is the eigenbasis of $H_0$ and the normalization is given by the canonical partition function $Z=\sum_k\exp{-E_k / k_B T}$.

We simulate readout by post-processing the final state $\rho_f$: From the density matrix, represented in the dressed basis, we obtain a vector of
populations $\vec{p}=\qty(p_k)$ by taking the absolute square of the diagonal. This is consistent with
a slow (or weak) readout scheme in experiment.
Measurement and classification errors are modeled with a misclassification (confusion) matrix
$(p_{i \rightarrow j})_{ij}$ \cite{stehman1997selecting} such that the measured populations are
\begin{equation}\label{eq:confusion}
\widetilde{p}_j=\sum_i p_{i \rightarrow j} p_i \ .
\end{equation}
To simulate an experimental measurement with an average of $l$ repetitions, we draw from a multinomial distribution of $l$ trails and with probabilities $\widetilde{p}_j$.

\subsection{Optimizations}
For open and closed-loop optimal control as well as model learning, performing optimization processes is required. 

\subsubsection{Open-loop Model-based Control: \cOne}\label{C1}
In the typical setting of open-loop optimal control \cite{Glaser2015,Machnes2011}, given a model of
a system, we search for the optimal control pulses to drive the system to a desired state
or generate a certain gate. 
Pulses are parameterized by an analytic ansatz (e.g. Gaussian pulse with DRAG correction \cite{DRAG} to remove
Fourier components coupling to leakage levels), or by direct AWG samples. Constraints may be imposed to conform
with experimental feasibility, such as power and bandwidth limitations. The goal function to be minimized is
selected depending on the specified optimal control task, e.g. state infidelity for state transfer
problems, or unitary trace infidelity for quantum gates \cite{emerson-avfid, Machnes2011}. 
We suggest the use of average gate infidelity as the goal function, as it is experimentally accessible by 
performing RB or QPT, allowing comparison of performance in simulation and experiment.

Formally, the controls are parameterized as a vector of real numbers $\alpha$. Given a goal
function $f_1(\alpha)$, we search for $\min_{\alpha}f_1(\alpha)$.
Optimal control methods such as GRAPE \cite{GRAPE}, Krotov \cite{Krotov1, Krotov2, Koch-Krotov-Main,
	KrotovPython}, and GOAT \cite{Machnes2018} have been devised to determine the gradient
$\partial_{\alpha} f_1(\alpha)$ in order to facilitate convergence. 
These methods require a specific formulation of the problem and the analytical calculation of the gradient any additional elements in the model, whereas in \CCC, automatic differentiation allows to systematically account for any
model feature, including, for example, line response functions or SPAM error.

The disadvantage of gradient-based algorithms is their propensity to get trapped in local minima.
The severity of the problem is reduced by using a hierarchy of progressively more complex control
ans\"atze. If this is insufficient, a short preliminary gradient-free search to find the
convergence basin most often resolves the problem.

\subsubsection{Closed-loop Model-free Calibration: \cTwo}\label{C2}

In calibration, a given pulse is optimized to improve a figure of merit
$f_2(\alpha)$, computed from experimental measurement results. 
In addition to gradient-free optimization algorithms, there are methods to
approximate the gradients (e.g. \cite{leng2019robust}), however, such approaches are generally less efficient
than gradient-free algorithms \cite{Machnes2018, nocedal2006no}
as they require a high number of evaluations \cite{feng2018gradient}. 
If the initial point of the optimization is given by \cOne, this implements the already established
Ad-HOC \cite{Egger2014} method.
During calibration, sets of control parameters $\alpha_j$ are sent to the experimental setup,
alongside instructions of how to evaluate the current controls. 
For evaluating gate-sets, we suggest the ORBIT figure of merit, as it naturally performs a twirling of all sources of error, providing a single number to optimize. 
However, protocols tailored to specific needs can also be used, e.g. to obtain a desired conditional phase \cite{collodo2020implementation}.
\cTwo\ then optimizes the control parameters $\alpha_j$ to minimize a figure of merit.

While specialized measurements provide a straightforward way to fine-tune controls related to
specific device properties, they do not generally account for interdependency.  For more complex
setups with many parameters, such calibrations cannot be done without extraordinary effort
\cite{kelly2018physical}. In contrast, \CCC\ employs modern gradient-free optimization algorithms,
such as CMA-ES (see App. \ref{supp:grad-free} for further discussion), capable of optimizing 
dozens of parameters simultaneously, automating the task.


\subsubsection{Model Learning: \cThree}\label{C3}

Extracting the model from a data-set $\mathcal{D}$ can be thought of formally as analogous to the \cOne\
optimization task, where one varies the \emph{model} parameters instead of the control parameters.
For each measurement outcome $m_k$ in the data-set,
\begin{equation}
\mathcal{D}=\{S_k\mapsto m_k\}_j\, ,
\end{equation}
the corresponding gate or pulse sequences $S_k(\alpha_j)$ with control parameters $\alpha_j$ are used to
simulate the model's prediction $\widetilde{m}_{j,k}=\widetilde{m}\qty(S_k(\alpha_j), \beta)$.
The model learning goal function
\begin{equation}
f_3\qty(\mathcal{D} \vert \beta) = f_3\qty(\qty{\widetilde{m}_k(\beta)},  \qty{m_k})
\end{equation}
quantifies the match between the data-set and the simulation of a system with
parameters $\beta$. In this paper, we use a rescaled log-likelihood 
\begin{equation}
f_\text{LL}\qty(\mathcal{D} \vert \beta) =
\frac{1}{2K}\sum_{k=1}^K\qty(\qty(\frac{m_k-\widetilde{m}_k}{\widetilde{\sigma}_k})^2-1)\ ,
\end{equation}
where the $\widetilde{\sigma}_k$ is the standard deviation of a binomial distribution with mean
$\widetilde{m}_k$, resulting in a variation of the Mahalanobis distance \cite{mahalanobis1936generalized}.
This function is strictly correct under the Gaussian assumption and a two-level readout.
See Sec. \ref{sec:f_LL} for the extension for a multiple outcome readout. 
The measurement process on any physical device is noisy, i.e. each $m_k$ is an estimate of a true underlying $\mu_k$.
Therefore, a realistic data-set $\mathcal{D}$ cannot be matched exactly by a deterministic simulation.
The function $f_\text{LL}$ is designed such that, for $n$ data points, its expectation value is $0$ when the
model predicts the means $\mu_k$ correctly, and $\frac{1}{2}n^2$ if the distance is $\mu_k - \widetilde{m}_k= n\sigma_k$
for all $k$s, according to Eq.  (\ref{eq:hLL_semi_mean}). 
This provides a more intuitive measure of model match than the
abstract value of $f_\text{LL}$, i.e. it allows to make a statement like ``the model differs from the experiment by approximately $n$
standard deviations''.

Due to the complexity of the physical systems, a potentially high number of interdependent parameters and complex features of the landscape, it is difficult for the optimization to converge to the global optimum. 
Therefore, we take the tried-and-tested experimental approach of
starting with a simple model and iteratively refining it.
We modify the model and repeat the \cThree\ fit, optionally retaining the
optimized parameters which are shared by the previous and new model.
Alternatively, we collect additional data and repeat the optimization on the same model.
We emphasize that at each of these steps the physicists' insights are required to evaluate the
optimization's results, extend or discard models and decide whether collecting additional data is required.
Furthermore, employing a gradient-based algorithm can, depending on the initial point, result in a 
local minimum. The optimizations
presented here were successful when starting with a gradient-free CMA-ES search,
known to be robust against local minima, switching over to the faster converging gradient-based L-BFGS method when a promising parameter region is identified. However, further research is required to find the best optimization strategy.

Outside the countless parameter-specific measurements, there are many approaches that aim to automate or generalize the task of characterization, such as: Bayesian learning (with Hamiltonian description 
\cite{schirmer2010quantum,stenberg2014efficient,stenberg2016characterization,stenberg2016simultaneous,sergeevich2011characterization,wiebe2014hamiltonian,wiebe2014quantum,wiebe2015quantum,wang2017experimental,Cole2005}, and more general-purpose \cite{granade2012robust, Krenn2016, melnikov2018active, krenn2020computerinspired, gentile2020learning,evans2019scalable,ralph2017multiparameter})
, system-identification \cite{kosut2002identification,schirmer2004experimental,oi2012quantum,burgarth2012quantum,zhang2015identification,hou2017experimental}, compressed-sensing \cite{shabani2011estimation,rudinger2015compressed,wiebe2015quantum}, neural network \cite{lusch2018deep,flurin2020using,dalgaard2020global}, and others \cite{krastanov2019stochastic}.

In contrast, we note that in \CCC\ a model takes explicit values for all its parameters, and is \emph{not}
represented as a high-dimensional distribution over model parameter space. This choice is driven by
classical computation-load considerations: Because the \CCC\ model is
highly detailed, and, as consequence, associated simulations are non-trivial, we believe a full
Bayesian approach to any of \CCC\ optimizations is not computationally viable at this time. 

\subsection{\cThree\ Model Fitting Goal Function}\label{sec:f_LL}

When performing a series of experiments, $k\in\qty[1,\ldots,K]$, on a quantum device, each experiment $k$ is repeated a number of times and the normalized occurrences of the measurement outcomes are store in a result vector $m_k$. These are collected in $\left\{ m_{k}\right\} _{k=1,\ldots,K}$, or $\qty{ m_{k}} $ for shorthand.
Given a model and its parameters $\beta$, we aim to quantify how likely it is that it underlies the observed data with a function $f_3\qty(\qty{m_{k}}\vert\beta)$.
Hence, we need to determine the distance between the experimental result, $m_{k}$, 
and the model prediction, $\widetilde{m}_{k}(\beta)$. We define $p_{k}(m_{k}\vert\beta)$
be the model-predicted probability distribution function (PDF) for the result of experiment $k$. As the $m_{k}$ are sampled from the readout distribution, we do not expect $m_{k}=\widetilde{m}_{k}(\beta)$. 
Rather, we aim to define the function $f$ such that its expectation value, $\textrm{E}\qty[f\qty(\left\{ m_{k}\right\},\beta)]$, is zero if the underlying distributions from which the $\qty{m_{k}}$ are drawn are the same as the model-predicted PDFs.

\subsubsection{The Gaussian Assumption}
To simplify calculation of $\textrm{E}\qty[f\qty(\left\{ m_{k}\right\},\beta)]$, we can make some assumptions regarding
the underlying distributions. The natural $p_{k}\qty(m_{k}\vert\beta)$ PDF is multinomial, determined by the dimension of the qubit Hilbert space $d_k$ (or 
binomial if dealing with a single qubit with no leakage levels). Under the assumption that for a large 
number of shots all possible readouts values are likely to appear, then by the central limit theorem (De Moivre--Laplace theorem), we can approximate $p_{k}$ with a multivariate normal distribution. 
Although, the multinomial distribution has a non-diagonal covariance matrix, one can diagonalize the distribution and decompose it as a product of one-dimensional Gaussian distributions.
Thus, we write the PDF as a sum of $\sum_{k}(d_{k}-1)$ such distributions, redefine $K$ to equal the previous $\sum_k(d_{k}-1)$ and the $\qty{m_{k}}$ to be their means.

\subsubsection{The model match distribution}

We shall use $\left\{ \widetilde{\mu}_k(\beta)\right\} $ and $\left\{ \widetilde{\sigma}_k(\beta)\right\} $
to denote the mean and standard deviation of the model-predicted PDFs (after Gaussian assumption and multinomial
diagonalization), and $\left\{ \mu_k\right\} $ and $\left\{ \sigma_k\right\} $ to denote the commensurate
experimental values.
We note that $\left\{ \mu_k\right\} $ and $\left\{ \sigma_k\right\} $ are unknown and unmeasured, and
$\left\{ m_k\right\} $ only provides an estimate of the mean. 
The simulation values $\left\{ \widetilde{m}_k\right\} $ on the other hand are deterministic and thus represent an exact estimate of the mean, hence 
$\left\{ \widetilde{m}_k\equiv\widetilde{\mu}_k\right\}$.

The model-predicted PDF is given by the product of normalized Gaussian distributions, and gives the 
likelihood of the $\left\{ m_k\right\}$ given the model parameters $\beta$ as
\begin{equation}
 p\left(\left\{ m_k\right\} \vert\beta\right)
 =\prod_{k}p\left(m_k \vert\beta\right)\ \qq*{, where }
 \end{equation}
 \begin{equation}
 p\left(m_k \vert\beta\right) = \frac{1}{\sqrt{2\pi}\widetilde{\sigma}_k}\exp\left(-\frac{1}{2}\left(\frac{m_k-\widetilde{\mu}_k}{\widetilde{\sigma}_{k}}\right)^{2}\right)
\end{equation}
are the individual Gaussian distributions. We then construct the goal function as the average log-likelihood, rescaled to give the desired expectation value,
\begin{equation}
	\begin{aligned}
	f_{\textrm{LL}}\left(\left\{ m_{k}\right\} \vert\beta\right)
	&=-\log\qty[\sqrt[K]{p\left(\left\{ m_k\right\} \vert\beta\right){\prod_{k}\sqrt{2\pi}\widetilde{\sigma}_k}\sqrt{e}}]\\
    &=\frac{1}{K}\sum_k\frac{1}{2}\left(\left(\frac{m_k-\widetilde{\mu}_k}{\widetilde{\sigma}_k}\right)^{2}-1\right)\ .
	\end{aligned}
\end{equation}

Here the $\sqrt[K]{\cdot}$ gives the average of the log-likehoods, 
the $\sqrt{2\pi}\widetilde{\sigma}_k$ removes the normalization of the Gaussians, such that they take value 1 
when $m_k-\widetilde{\mu}_k=0$, and the log-likelihood is zero,
and the $\sqrt{e}$ removes the residual part of the expectation caused by the noise in the $\left\{ m_k\right\}$.

Then, in the general case, when the Gaussians determined by the model are not the same as the Gaussians in the experimental data:

\begin{equation}
\begin{aligned}
    &\textrm{E}\left[f_{\textrm{LL}}\left(\left\{ m_{k}\right\} \vert\beta\right)\right]\\
    &=
    \frac{1}{2K}\sum_k\left(\left(\frac{\mu_k-\widetilde{\mu}_k}{\widetilde{\sigma}_k}\right)^{2}+\left(\frac{\widetilde{\sigma}_k}{\sigma_k}\right)^{2}-1\right) \ ,
\end{aligned}
\end{equation}
\begin{equation}
\begin{aligned}
   & \textrm{Var}\left[f_{\textrm{LL}}\left(\left\{ m_{k}\right\} \vert\beta\right)\right]\\
   & =
    \frac{1}{K^{2}}\sum_{k}\left(\frac{\sigma_k}{\widetilde{\sigma}_k}\right)^{2}\left(\left(\frac{\mu_k-\widetilde{\mu}_k}{\widetilde{\sigma}_k}\right)^{2}+\frac{1}{2}\left(\frac{\sigma_k}{\widetilde{\sigma}_k}\right)^{2}\right) \ .
    \end{aligned}
\end{equation}
In the limit that both distributions have the same standard deviation  $\sigma=\widetilde{\sigma}$
\begin{equation}
    \textrm{E}\left[f_{\textrm{LL}}^{\sigma\leftarrow\widetilde{\sigma}}\left(\left\{ m_{k}\right\} \vert\beta\right)\right]=\frac{1}{2K}\sum_k\left(\frac{\mu_k-\widetilde{\mu}_k}{\widetilde{\sigma}_k}\right)^{2}
    \label{eq:hLL_semi_mean}
    \end{equation}
\begin{equation}
    \textrm{Var}\left[f_{\textrm{LL}}^{\sigma\leftarrow\widetilde{\sigma}}\left(\left\{ m_{k}\right\} \vert\beta\right)\right]=\frac{1}{2K}+\frac{1}{K^{2}}\sum_{k}\left(\frac{\mu_k-\widetilde{\mu}_k}{\widetilde{\sigma}_k}\right)^{2} \ .
\end{equation}

Equation (\ref{eq:hLL_semi_mean}) then represents the square of the Mahalanobis distance (standardized Euclidean distance), giving an intuitive way to
scale the $f_{\rm LL}$ function to understand the model match score. Indeed, when the model is exact
and $\mu_k = \widetilde{\mu}_k$ we get
$\textrm{E}\left[f_{\textrm{LL}}^{\textrm{exact}}\left(\left\{ m_{k}\right\} \vert\beta\right)\right]=0$.
We note, however, that the function can take values below 0 as the variance for the exact case is 
$\textrm{Var}\left[f_{\textrm{LL}}^{\textrm{exact}}\left(\left\{ m_{k}\right\}\vert\beta\right)\right]=\frac{1}{2K}$.
Such values indicate the standard deviation expected by the model, $\widetilde{\sigma}_k$ is larger than the standard deviation observed experimentally, ${\sigma}_k$.


\subsection{Model analysis}
Both during and after the learning process, it is beneficial to interrogate the model to estimate
its properties and their impact on the system behavior. As part of the \CCC\ tool-set we perform 
sensitivity analysis for system parameters: Sweeping a single parameter, e.g. qubit frequency,
across the range of interest, while keeping other parameters at their current best value, evaluating 
the model match score at each point, as seen in the example (Fig. \ref{fig:sens-freq}).
The result is a 1-D cut through the optimization landscape that may exhibit a well-defined minimum,
multiple extrema indicating a difficult optimization, or even appear flat in the case when
a parameter does not affect the behavior of the current experiment. This landscape depends on both the
selected model and data it is compared to.
Depending on the ruggedness of the sensitivity, one might choose to utilize a gradient-based algorithm
from the start or to first perform a gradient-free exploratory search to avoid local minima.
In the case of a flat sensitivity, there are two courses of action: If the parameter
is of little interest for successive experiments, it may be removed or set to a convenient value within the
flat range; otherwise, one needs to design an experiment producing additional data that is sensitive to
the parameter. The physicists' knowledge of common experimental practices (e.g. Rabi, Ramsey, Hahn echo sequences) and intuition guides the decision for the experiment design.
When suspecting correlations between parameters, cuts in single dimensions are not enough and higher 
dimensional sweeps are necessary. After a successful learning process, the sensitivity analysis gives 
an estimate of the precision to which each parameter has been determined.

Furthermore, the simulation allows insight into the behavior of the system.
Using well established methods such as time-resolved state and process tomography, it is possible to identify 
coherent errors, such as leakage out of the computational subspace, over-rotations, and the effects of noise. 
A Good Model also provides the basis for an error budget, as it contains the same limitations as
the experiment it accurately predicts.
The model can be used for extrapolation by idealizing certain aspects suspected as causes of
infidelity (e.g. $T_1$ setting to infinity), and re-deriving control pulses using a \cOne\
optimization. The respective gain in fidelity gives an estimate of the error that this aspect is responsible for,
suggesting areas of improvement for future devices. 


\section{Discussion and Outlook}\label{sec:discussion}

In conclusion, we have described \CCC, an integrated methodology to improve quantum device
performance that combines characterization, calibration and control. We have detailed its approach and
implementation, demonstrating, on a synthetic QPU device, the individual methods and how they are synthesized into a more integrated concept.
Analyzing single-qubit calibration data we successfully extracted an accurate model of the device,
including realistic experimental considerations: line transfer functions, 
limitations of control electronics, readout error and finite operating temperature. 
From this model we were able to derive a working high-fidelity two-qubit gate, without requiring any further calibration.

This approach represents a holistic theoretical take on the experimental workflow of a complex
quantum computing experiment, that takes into account interactions between different tasks of an
experimental lab.  \CCC\ provides a path to achieve, starting from an incomplete understanding of
the system, both high-fidelity pulses and an accurate model. It integrates the tasks of open-loop
control (that would require an already accurate model) and of calibration (that would require
an experiment-specific fine-tuning procedure).  Most notably, it provides the tools to reflect on the experiment
outcome and gate performance, improving the model description of the system and providing insight
into its behavior.
\CCC\ is not a ``black-box'' experiment controller that replaces physicists or engineers -- rather,
it reduces tedious tasks allowing for interaction with the quantum device on a more structural level.  
Instead of simply producing high-fidelity operations, \CCC\ provides meaningful output in the form
of a Good Model of the system, and other insights such as an error budget and a sensitivity analysis.
In this sense, \CCC\ is not to be confused with any single optimal control or benchmarking technique,
as it includes results from decades of research in these fields aimed at making controls that allow
to actually reach high fidelities efficiently \cite{Glaser2015,wilhelm2020introduction}, unifying
them into one framework.

We expect that the application of \CCC\ will first benefit scalable implementations of quantum
processors based on manufactured solid-state systems, such as superconducting and semiconducting
qubits. There, the dependence of model parameters on fabrication means that many elements of the
model are in fact uncertain. On the other hand, other scalable implementations of quantum computers
contain such elements directly in their quantum description: Ion trap gates involve degrees of
freedom of the trap, impurity spins depend on their detailed position etc. -- thus we expect that
\CCC\ will also play a key role in those types of systems.

In the near future, we intend on extending the initial version of the \CCC\ tool-set
to the generation of robust controls, automatic experiment design, multi-parameter sensitivity analysis,
active model learning, and more. The simulator will be enhanced to include non-Markovian noise, a detailed
simulation of the readout process, echoes on control lines and all other phenomena needed to produce a Good
Model of real-world systems. \CCC\ is currently being integrated with existing quantum computing software 
stacks, which would allow users to study custom pulse schedules \cite{Mckay2018, Alexander2020} and perform 
model learning based on data gathered from quantum computers, for example, with Qiskit Ignis \cite{Qiskit}. 
Experimental application of \CCC\ is ongoing (e.g. \cite{Werninghaus2021}).

It is our hope that \CCC\ will not only provide insights into, and assist optimization of, current experiments, but also help guide the design of next-generation quantum devices, be it manually or by integration into automatic hardware design frameworks \cite{Krenn2016,melnikov2018active,macleod2020self}.


\section*{Acknowledgements}

This work was supported by the European Commission through the Marie Curie ETN QuSCo (Grant Nr. 765267) and the OpenSuperQ project (Grant Nr. 820363), by the Intelligence Advanced Research Projects Activity (IARPA) through the LogiQ (Grant Nr. W911NF-16-1-0114) and by the Germany Ministry of Science and Education (BMBF) through project VERTICONS (Grant Nr. 13N14872).
\appendix


\section{Survey of parameter specific characterization}

The task of characterization of quantum devices in general has received
extensive attention. It would be presumptuous of us to even attempt
a complete survey, therefore, we shall limit ourselves to a very
limited look at a subset of model-specific methods we have subjectively 
found informative to our own work.

The standard approach at addressing the lack of a Good Model, as defined above, is to perform a long list of model-specific characterization experiments, each designed to measure a different parameter of the model: measure parameters of the readout resonator using frequency sweeps; qubit frequency measurements and relaxation time $T_1$ require Rabi experiments \cite{rabi1937space} (and with some extra effort the higher levels can be extracted); Ramsey \cite{ramsey1950molecular} and Hahn echo measurements \cite{hahn1950spin} provide dephasing data (under the Markovian assumption, which is known to be an over-simplification \cite{pokharel2018demonstration,ferrie2018bayesian}); measuring the control line response functions \cite{rol2020time,van2019impact,jerger2019situ,hincks2015controlling,hincks2015controlling,gustavsson2013improving}, the noise spectra \cite{ferrie2018bayesian, gupta2019adaptive, harper2019efficient,ball2020software}, continuous drifts in system parameters \cite{kelly2016scalable,casparis2016,miquel1997quantum,bialczak20071}, and discontinuous jumps in parameters such as $T_1$ \cite{burnett2019decoherence,niu2019learning}; state Preparation and Measurement Errors (SPAM) can be extracted from Randomized Benchmarking (RB; e.g. \cite{Knill09}) or dedicated procedures, such as \cite{zhang2020experimental}; qubit cross-talk can be measured by the method described in \cite{mundada2019suppression,murali2020software} and many more. 
Model specific methods also exist for learning spin chain, lattice Hamiltonians and other multi-particle systems with a predefined network topology under limited access \cite{burgarth2009coupling,di2009hamiltonian,de2016estimation,sone2017hamiltonian,yang2020complete}.


\section{Choice of Gradient-Free Algorithms}\label{supp:grad-free}
\begin{figure*}[t]
	\centering
	\includegraphics[width=\linewidth]{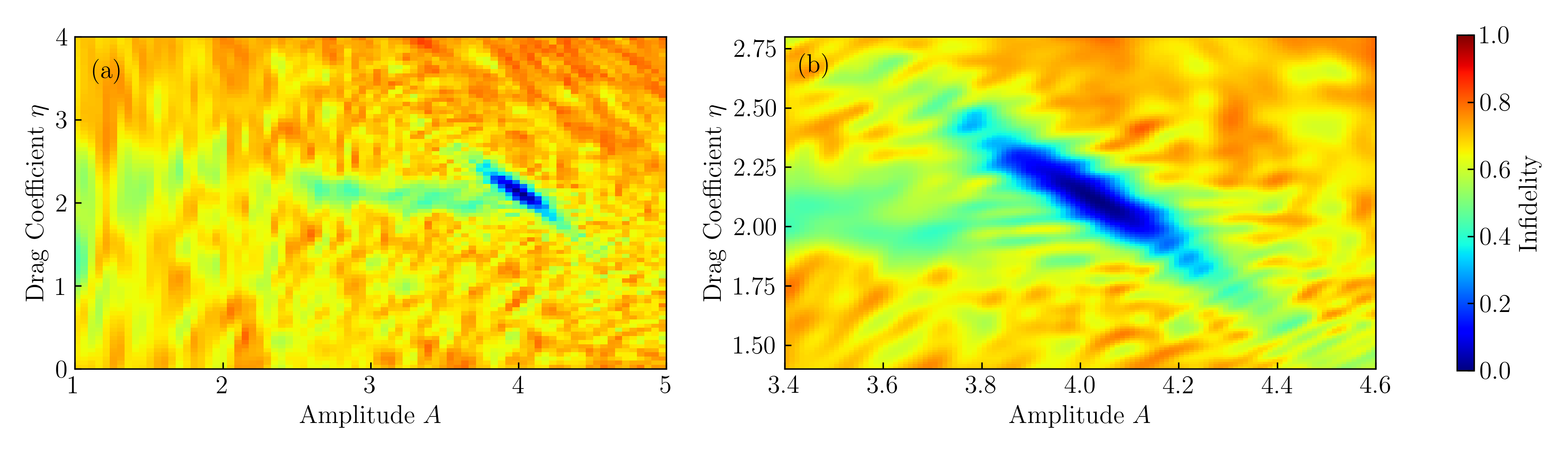}
	\caption{
		A 2D cut through the landscape of a calibration goal function. The system chosen is a
		multi-level qubit, with the control sequence being a single length RB-sequence. Control
		parameters $\alpha_j$ were the Gaussian pulse amplitude, DRAG coefficient and
		frequency offset of a DRAG-corrected pulse. The RB-sequence
		was chosen to implement the identity operation, ideally leaving the qubit in the
		ground-state. The infidelity (seen as the color ranging from 0 to 1) is defined as the
		overlap of the final state with the system's ground state.
		(a) The landscape of the simulated system, as a cut through the plane of pulse
		amplitude $A$ and DRAG coefficient $\eta$, for a fixed RB-sequence. Multiple local minima can be
		observed.
		(b) A higher resolution plot of the same landscape. Further local minima can be
		observed in the neighborhood of the global minimum.
	}
	\label{fig:rb-sequence-landscape}
\end{figure*}
Naturally, using an automated approach to solve the gradient-free optimization problem of calibration seems promising, and many
experiments are currently using the \textit{Nelder-Mead algorithm}
\cite{Nelder-Mead,Nelder-Mead-slow-dim} for their calibration tasks. As an example, Fig. \ref{fig:rb-sequence-landscape} shows an underlying parameter landscape of an optimization, obtained by simulating DRAG corrected pulses. Calibration can be rather difficult, as depending on the starting position the optimizer, algorithms will have to overcome local minima and deal with intrinsic noise. It is further noteworthy that this landscape is rather unique to the used parametrization, the chosen goal function and ultimately the properties of the physical system.

Research of gradient-free optimizers is a vast and active field, with hundreds of published black-box optimizers, including Evolutionary Strategies 
(ES), Particle Swarm Optimization (PSO), Differential Evolution (DE), Random Search, Simultaneous 
Pertubation Stochastics Approximation (SPSA), Nelder-Mead Method, Bayesian Optimization, and more. From a preliminary investigation, we recommend evolutionary strategies such as CMA-ES, which is currently the default in \cTwo. This algorithm performed well in most cases, exhibited good robustness to noise, can handle local extrema and requires relatively few evaluations. Similar conclusions have been reached independently by \cite{rapin2019}.
We make no claim as to the optimality of the optimizer chosen, and defer
a more detailed discussion of the subject to future publications.


\section{Open-source implementation}\label{supp:open-source}

\CCC\ is implemented as an open source project available at \url{https://q-optimize.org} under the
Apache 2.0 license. The software is written in Python to interface conveniently with common experiment
controllers, and has already been used in tandem with PycQED \cite{pycqed}, Labber \cite{labber} and LabView \cite{kalkman1995labview}. 
The interface can occur at various levels of abstraction, from sharing control parameters to sampled waveform values.
A modular design allows for Hamiltonian or Lindbladian descriptions of common physical systems (fixed and flux-tunable
qubits, resonators, different types of coupling), specification of a list of devices to model the signal chain of the
experiment (local oscillator, AWG, mixers, distortions and attenuations), different types of readout processing, and
various fidelity functions.
All components can be edited by the user or taken from reference libraries, accommodating to different needs. 
Configurations and data are stored as JSON files, and the full capabilities are accessible as command-line scripts, allowing for easy automation.

Numeric calculations are performed using TensorFlow \cite{tensorflow2015-whitepaper}: The simulation 
of the dynamics and the pre and post processing are formulated as a network, with well-defined
inputs (e.g. control and model parameters) and outputs (goal function values), connected by many nodes, each performing a relatively simple operation (e.g. matrix exponentiation).
TensorFlow enables the numerical computation of the Jacobian of a calculation -- 
the gradient of each of the network outputs with respect to the network inputs (this
capability is the evolution of what is known as back-propagation learning process in neural networks
\cite{rumelhart1986learning}).
This process of automatic differentiation facilitates the modular structure, as any new component inherits
this property, removing the need to analytically derive its gradient.
Furthermore, the TensorFlow simulator is scalable, allowing deployment on a variety of high-performance computing hardware.

We note prior efforts simulating quantum circuits which allow for automatic differentiation, e.g.  \cite{leung2017speedup,abdelhafez2019gradient}, as well as large-scale simulations of quantum circuits, e.g. \cite{willsch2017gate,willsch2018testing,willsch2020supercomputer}.

Each component of the control stack and model needs to conform to a general boilerplate
that specifies what parameters it contains and how they are used.
In this modular design, each class represents a component of the experiment that takes an
input applies some parameter-dependent function to it and returns a result. 
For example, an envelope function for pulses would have this structure:
\lstset{
	basicstyle=\footnotesize\ttfamily, frame=tb, language=python
}
\begin{lstlisting}[linewidth=\columnwidth,breaklines=true]
import tensorflow as tf
...
def my_envelope_fuction(t, parameters):
    amplitude = parameters["amp"]
    p2 = parameters["p2"]
...
return tf.some_math_function(amplitude, p2, t)
\end{lstlisting}
The only requirement to this code is that mathematical functions have to be taken from the TensorFlow package to allow for
automatic differentiation.
As an example of a control stack element, the finite rise time of an AWG is realized with the following code:
\begin{lstlisting}[linewidth=\columnwidth,breaklines=true]
class Response(Device):
def __init__(..., rise_time, ...):
    ...
    self.params['rise_time'] = rise_time
    
    def process(self, iq_signal):
	    ...
	    t = self.params['rise_time']
	    sigma = t / 4
	    ...
	    # Convolution with a Gaussian
	    ...
	return signal
\end{lstlisting}
A signal processing chain is represented by putting the output of one control stack element into the next.
In calculating figures of merit, the user can choose from a library of functions or supply their own.
For example, the infidelity of a state transfer process from $\ket{\psi_0}$ to $\ket{\psi_\text{ideal}}$, 
implemented by the simulated propagator $U$ as follows:
\begin{lstlisting}[linewidth=\columnwidth,breaklines=true]
def state_transfer_infid(U, psi_ideal, psi_0):
    psi_actual = tf.matmul(U, psi_0)
    overlap = tf_abs(
    tf.matmul(
        tf.linalg.adjoint(psi_ideal),
        psi_actual
        )
    )
    infid = 1 - overlap
return infid
\end{lstlisting}

\end{document}